\documentclass[aps,pre,showpacs,showkeys,twocolumn,superscriptaddress,
  notitlepage, floats,floatfix,preprintnumbers,longbibliography]{revtex4-1}
\usepackage{amssymb,amsmath}
\usepackage{bm,url,graphics,subfigure,epsfig,rotating,float}
\usepackage{mathrsfs} 
\usepackage{color,soul,graphicx,dcolumn,enumitem}
\usepackage{ulem}
\usepackage{hyperref}
\usepackage{euler}
\graphicspath{{./}}

\def \Psic {\Psi_{\rm crit}}
\def \atwo {a_{\rm 2}}
\def \ath {a_{\rm 3}}
\def \af {a_{\rm 4}}
\def \RC {{\tt Run C }}
\def \RS {{\tt Run S }}
\def \RMo {{\tt Run M1 }}
\def \RMt {{\tt Run M2 }}
\def \RMth{{\tt Run M3 }}
\def \Tast {T_{\ast}}

\def \MSD  {\avg{R^2}}
\def \fext {f^{\rm ext}}
\def \ffext {\bm{f}^{\rm ext}}

\def \xvec {{\bm{x}}}
\def \fvec {{\bm{f}}}
\def \kvec {{\bm{k}}}

\def \fvec {{\bm f}}

\def \Vvec {{\bm V}}

\def \urms {u_{\rm rms}}

\def \mD {\mathcal{D}}

\def \us {\bm{u}_{\rm s}}

\def \zbulk {\zeta_{\rm bulk}}
\def \deli {\partial_{i}}
\def \delj {\partial_{j}}
\def \delk {\partial_{k}}
\def \ui   {u_{i}}
\def \uj   {u_{j}}
\def \uk   {u_{k}}

\def \LI {L_{\rm I}}

\def \zhat {\hat{z}}

\def \rz  {r_{\rm 0}}

\def \km {k_{\rm max}}

\def \uu  {\bm{u}}

\def \ff  {{\bm f}}

\def  \xx  {{\bm x}}

\def  \XX  {{\bm X}}
\def \XXm  {\bm{X}^{\rm m}}

\def  \VV  {{\bm V}}

\def \cs {c_{\rm s}}

\def  \RR  {{\bm R}}

\def \grad {{\bm \nabla}}
\def \curl {{\bm \nabla} \times}
\def \dive {{\bm \nabla}\cdot}

\def \delt {\partial_t}

\newcommand{\avg}[1]{\left\langle #1\right\rangle}

\def \Rey  {\mbox{Re}}

\def \Ma  {\mbox{Ma}}

\def \TL {T_{\rm L}}

\def \kf  {k_{\rm f}}

\def \urms  {u_{\rm rms}}

\def \Np  {N_{\rm p}}

\def\i{{\rm i}}

\def \CL {C_{\rm L}}
\def \TE {T_{\rm E}}

\def \tfp {t_{\rm fp}}

\newcommand{\eq}[1]{~(\ref{#1})}

\newcommand{\Fig}[1]{Fig.~(\ref{#1})}
\newcommand{\subfig}[2]{Fig.~(\ref{#1}#2)}

\newcommand{\bfig}{\begin{figure}}
\newcommand{\efig}{\end{figure}}
\newcommand{\bc}{\begin{center}}
\newcommand{\ec}{\end{center}}
\newcommand{\bea}{\begin{eqnarray}}
\newcommand{\eea}{\end{eqnarray}}

\newcommand{\SMat}[1]{Supplementary Material \cite{SM}}
\begin{document} 
\title{Shock-driven superdiffusion in isothermal compressible turbulence}
\author{Sadhitro De}
\email{sadhitro.de@physics.ox.ac.uk}
\affiliation{Department of Physics, University of Oxford, United Kingdom}
\author{Dhrubaditya Mitra}
\email{dhrubaditya.mitra@su.se}
\affiliation{Nordita, KTH Royal Institute of Technology and
Stockholm University, Roslagstullsbacken 23, 10691 Stockholm, Sweden}
\altaffiliation{Digital Futures, Stockholm, Sweden}
\author{Rahul Pandit}
\email{rahul@iisc.ac.in}
\affiliation{Department of Physics, Indian Institute of Science, Bangalore 560012, India}

\begin{abstract}
  In statistically homogeneous and isotropic incompressible turbulence,
  Taylor showed that the mean-square displacement (MSD) of Lagrangian tracers
  displays  characteristic diffusive behavior at late times.
  How does this important result change if we consider isothermal,
  compressible turbulence?
  We combine direct numerical simulations in two
  dimensions and theoretical arguments to uncover a temporal window in which
  the MSD is \textit{superdiffusive}.
  For time larger than a crossover time, MSD becomes diffusive.
  The crossover time increases as we decrease the soledoinality of the flow
  and can be larger than the large-eddy-turnoever time. 
  We develop a large-deviation approach that yields a natural rationalization
  of the superdiffusive behavior.
  Furthermore, we calculate the probability distribution function  for the
  tracer first-passage  time and use it to demonstrate how this superdiffusion
  can be modeled via heterogeneous diffusion.
 \end{abstract}
%---------------------
\maketitle
%-------------------

%%%%%%%%%%%%%%%%%%%%%%%%%%%%%%%%%%%%%%%%%%%%%%%%
Turbulent mixing and transport are key drivers of uncertainty in our knowledge
of the nature of clouds in terrestrial atmospheres and also of those in
interstellar media (ISM) which serve as the birthplace of stars and
planetesimals. 
Our current understanding of turbulence is broadly based on the following
foundational results: Kolmogorov's theory of the turbulent energy
cascade~\citep{K41, frisch}, Richardson's law of pair
dispersion~\citep{richardson1926atmospheric}, and Taylor's theory of
single-particle dispersion~\cite{Taylor1922Diffusion}. 
Research on turbulence in the past half-a-century has focused on understanding
and extending these results.
Most engineering, laboratory, and  natural
terrestrial  flows  can
be reasonably approximated to be \textit{incompressible},
but flows in the interstellar matter comprising the ``the swirling nebulae
  and the condensing, revolving, and exploding stars and galaxies''
  are \textit{compressible}.
All of the aforementioned theories are 
applicable to incompressible flows.
\textit{How do they generalize to strongly compressible hydrodynamics?}.
While there has been substantial research on the adaptation of Kolmogorov's
theory of energy cascade to strongly compressible
turbulence~\cite{Mordecai2004Review,
   Elmegreen2004Review_ISM,Larson1981SF, Padoan2011SFR,
   Federrath2012SFR, Orkisz2017SF, Pan2011Planetesimal_Clust, Boldyrev2002SF_MC,
   Boldyrev2002SupersonicISM, Kritsuk2007ISM, Federrath2009FractalDensity,
   Federrath2008DensityPDF, Federrath2011Machno, Federrath2013Universality,
   Schmidt2008Universal}, exploration of the fundamental aspects of
Lagrangian transport and mixing in such systems, which is key to
parameterizing the rates of formation of stars and planetesimals, has been
relatively limited~\citep{Klessen2003Diffusion, Zhang2016Inertial,
  Zhang2018SinglePl}.
Recently, we have shown how to adapt Richardson's law
to compressible turbulence~\citep{de2024uncovering, de2026intermittency}. 
Here, we turn to the generalization of Taylor's theory to
compressible turbulence.

Taylor's theory~\cite{Taylor1922Diffusion} states that in
\textit{incompressible}, homogeneous, and
isotropic turbulence the mean-square displacement, $\MSD$, of Lagrangian
particles or tracers scales with time $t$ as
$\MSD \sim  t^{\gamma}$ with $\gamma = 2$ for  $t\ll\TL$ 
and $\gamma = 1 $  for $t \gg \TL$;
$\TL$ being the velocity autocorrelation time of the
tracers. 
That is, tracers diffuse like Brownian particles at late times with a
constant effective diffusivity.
Over the years, various experimental and numerical studies have established
the veracity of Taylor's results in both two- and
three-dimensions~\citep{Xia2013Diffusion,Xia2014Taylor,Mordant2004Diffusion,
  Muller2007DiffusionMHD,Busse2007DiffusionMHD,Akhilesh2020FPT,
  Kameke2011LagDispExp}. 

We first use direct numerical simulations (DNS) in
  two dimensions (2D) to show that in certain models of compressible
  turbulence, beyond the initial phase of ballistic motion,
  $\MSD \sim t^{\gamma}$ with $\gamma \approx 4/3$ --
  \textit{tracers exhibit superdiffusion}, before giving way to diffusion. 
  We then develop an understanding of this behavior through
  large deviation theory and heterogeneous diffusion. 

%-----------------------------------------------------
\begin{figure*}
    \centering
    \includegraphics[width=0.99\textwidth]{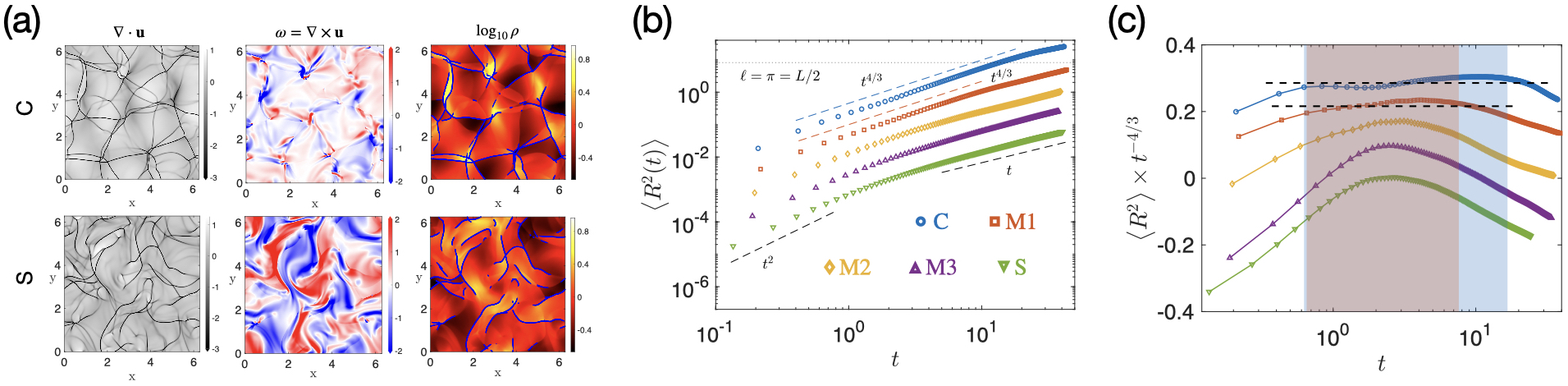}
    \caption{\textbf{(a)} Snapshots of $\dive\uu$, $\omega=\curl\uu$ and
      $\log_{10}\rho$ in \RC (top row) and \RS (bottom row);
      $\dive\uu$ and $\omega$ are normalized by their respective
      root-mean-square values. 
    Shocks are clearly visible as filamentary structures as $\dive\uu\ll0$. 
    Blue dots overlaid on the plots of $\log_{10}\rho$ are regions where
    $\dive\uu < -2\sqrt{\avg{|\dive\uu|^2}}$;
    $\rho$ is enhanced around these regions demonstrating clustering of
    tracers around shocks. 
     \textbf{(b)} 
     The mean-square displacement of tracers, $\MSD$, in the different runs
     versus time, $t$, on a log-log scale.
      The solenoidality ratios ($\Psi$) increases from top to bottom. 
      \textbf{(c)} The same data as in (b) with the vertical axis compensated
      with $t^{-4/3}$. The figure legends are the same as that in (b).
      Horizontal portions of the curves in this plot imply $\MSD\sim t^{4/3}$. 
      \RC and \RMo have distinct regimes of superdiffusion,
      $\MSD\sim t ^\gamma$ with $\gamma\simeq 4/3$,
      denoted by the blue and brown shaded regions, respectively . 
      This regime shrinks with an increase in $\Psi$ --
      it can be barely detected in \RMt and vanishes altogether in
      \RMth and \RS. 
      All curves in (b) and (c) are vertically shifted for the ease of
      visualization.
    }
    \label{fig:msd}
\end{figure*}
%-----------------------------------------%
Unlike incompressible turbulence, there is no unique model of
compressible turbulence. 
We perform direct numerical simulation (DNS) of compressible, isothermal,
ideal gas in 2D:
\begin{subequations}
\begin{align}
\delt \rho + &\delj(\rho\uj) = 0\,, \label{eq:cont}\\
\delt \ui + &\uj\delj\ui = -\frac{1}{\rho}\deli P \nonumber\\ 
&+ \frac{1}{\rho}\delj(\nu\rho S_{ij} + \rho\zbulk\delta_{ij}\delk\uk)
+ \fext_i\,. \label{eq:mom}
\end{align}
\label{eq:cNS}
\end{subequations}
Here $\rho(\xx,t)$ and $\uu(\xx,t)$ are the density and velocity of the fluid,
respectively, at the position $\xx$ at time $t$, $\nu$ and $\zbulk$ are the
coefficients of shear and bulk viscosity respectively, (the latter is
non-uniform and enables the resolution of shocks
[see \ref{sec:numsim} and Refs.~\cite{vonneumann1950method,
    Haugen2022ClusteringInertial}]), $S_{ij}$ is the traceless rate-of-strain
tensor, and the pressure, $P$, is given by the
equation of state $P=\cs^2\rho$ in which $\cs$, the sound speed, is a constant. 
This model is appropriate for interstellar media (ISM) and large-scale
astrophysical objects such as molecular clouds where the gas is transparent to
radiation~\cite{Federrath2012SFR,Federrath2013Universality,Mordecai2004Review}. 
In these systems, turbulence is typically driven by supernova bursts or
stellar winds.
We approximate this by adding the external force $\ffext$,
which is random, white-in-time, and limited to large scales. 
After the system reaches a statistically stationary state
we introduce $\Np = 10^6$ tracers and track their
evolution. 
The details of the DNS are given in the Appendix~\ref{sec:numsim}. 

The compressibility is parametrized by two dimensionless numbers: 
the turbulent Mach number,
$\Ma \equiv \urms/\cs$,
where $\urms$ is the root-mean-square flow velocity; and
the \textit{solenoidal ratio},
$\Psi= \avg{\rho|\us|^2}/\avg{\rho|\uu|^2}$ --
the ratio of the kinetic energy in the solenoidal component, $\us$, of the
flow velocity to the total kinetic energy
We perform five simulations with similar values of $Ma\gtrsim1$ and different
values of $\Psi$ by changing the relative energy injected into the solenoidal
and irrotational modes by the external force. 
They are called
\RC ($\Psi\approx 0.24$, \textit{irrotational forcing}),
\RMo ($\Psi\approx 0.35$),
\RMt ($\Psi\approx 0.46$),
\RMth ($\Psi\approx 0.61$) and
\RS ($\Psi\approx0.74$, \textit{solenoidal forcing}). 
The regime  $\Psi<1/2$ corresponds to predominantly irrotational
turbulence dominated by shocks typically found in  active star-forming
regions of molecular clouds~\cite{Mordecai2004Review, Federrath2012SFR,
  Federrath2008DensityPDF, Federrath2009FractalDensity,
  Federrath2013Universality, Orkisz2017SF}
We use the integral length scale, $\LI$, and the eddy-turnover time, $\TE=\LI/\urms$, as units of legth and time,
respectively. 
In all runs, shocks show up as one-dimensional manifolds characterized by
large negative values of $\dive\uu$ and $\rho$ is enhanced in their
neighbourhood This  indicates clustering of tracers around shocks
(see Fig.~\ref{fig:msd}(a)). 
Their typical lifetimes are significantly smaller than the duration of the
simulations.

Figure \ref{fig:msd}(b) shows a log-log plot versus time $t$
of the mean-square displacement $\MSD$, which is defined as
$\MSD=\avg{|\XX(t+0)-\XX(0)|^2}$, where $\XX$ denotes the instantaneous
position of a tracer particle, and $\avg{\cdot}$ represents averaging over
all tracers and different uncorrelated origin of time.
There are distinct regimes of \textit{superdiffusive transport},
$\MSD\sim t^\gamma$, for $t\gtrsim\mathcal{O}(1)$, in the predominantly
irrotational flows, \RC and \RMo.
This regime appears as horizontal regions in the compensated plots of
$\MSD\times t^{-4/3}$ in Fig.~\ref{fig:msd}(c). 
We note that, the superdiffusive regime is shorter in \RMo which has the higher
value of $\Psi$, as seen from the shaded regions in Fig.~\ref{fig:msd}(c). 
Within these regimes, the mean and standard deviation of the local slopes of
the log-log plots of $\MSD$ versus $t$ (\Fig{fig:locslop}) yield
$\gamma=1.34\pm0.07$ and $\gamma=1.34\pm0.06$ for \RC and \RMo, respectively. 
In \RMt, \RMth and \RS for which $\Psi\gtrsim1/2$, superdiffusion is absent
altogether -- only diffusion, $\MSD\sim t$, is observed at late times
in agreement with earlier results~\cite{Zhang2018SinglePl,
  Klessen2003Diffusion}. 
Therefore, we obtain the following generalization of Taylor's result to
compressible turbulence: for $\Psi<1/2$
\begin{equation}
     \MSD(t) \sim  \begin{cases}
     t^2 &\quad\text{for}\quad t\ll1\/,\\
     t^\gamma  &\quad\text{for}\quad \mathcal{O}(1)\lesssim t<\Tast(\Psi) \/,\\
     t &\quad\text{for}\quad t>\Tast(\Psi)\,,
     \end{cases}
     \label{eq:Taylor}
\end{equation}
where $\gamma \simeq 4/3$ and the crossover time,
$\Tast(\Psi)$, increases as $\Psi$ decreases.
This is the key result of our paper.
In the rest of the paper we elucidate the mechanism of
superdiffusion. 

As we consider homogeneous and isotropic flows,
$\MSD \approx 2 \avg{R^2_x}$ where $R_x$ is the $x$ component of the
displacement vector $\RR$. 
By definition
\begin{equation}
  R_x(T) = \int_0^{T} V_x(t) dt\/,
  \label{eq:Rx}
\end{equation}
where $\VV(t)$ is the Lagrangian velocity of the tracer.
Next we make the usual assumption that the Lagrangian velocity has a
maximum correlation time $\tau \ll T$.
This holds in our simulations, see \Fig{fig:VelCorrfn}. 
Hence, in the limit $T\to\infty$, the integral in \eq{eq:Rx} is
a sum of $N = T/\tau$ independent identically distributed (IID)
random numbers
$\xi = \int_0^{\tau} V_x(t)dt $.
By the large deviation principle~\citep{touchette2009large},
the probability distribution function (PDF), $P(\Delta)$,  of
$\Delta = R_x/T$ has the form
\begin{equation}
    \lim_{T\to\infty}\frac{1}{T}\ln P\left(\Delta=s\right) = -I(s)
    \label{eq:LDP}
\end{equation}
where the rate function $I(s)$ can be expanded in a power series
$I(s) = \atwo s^2 + \ath s^3 + \af s^4 +\cdots$.
Straightforward algebra shows that, if $I(s)$ is a quadratic,
i.e., $P(\Delta)$ is Gaussian in the limit $T\to \infty$,
then we obtain Taylor diffusion:
$\avg{R^2_x}\propto T$ for $T\to \infty$.
In other words, in the limit $T\to \infty$, there is
one unique dynamic exponent $z=2$ that relates
characteristic time scales to length scales. 

\begin{figure}
    \centering
    \includegraphics[width=0.99\columnwidth]{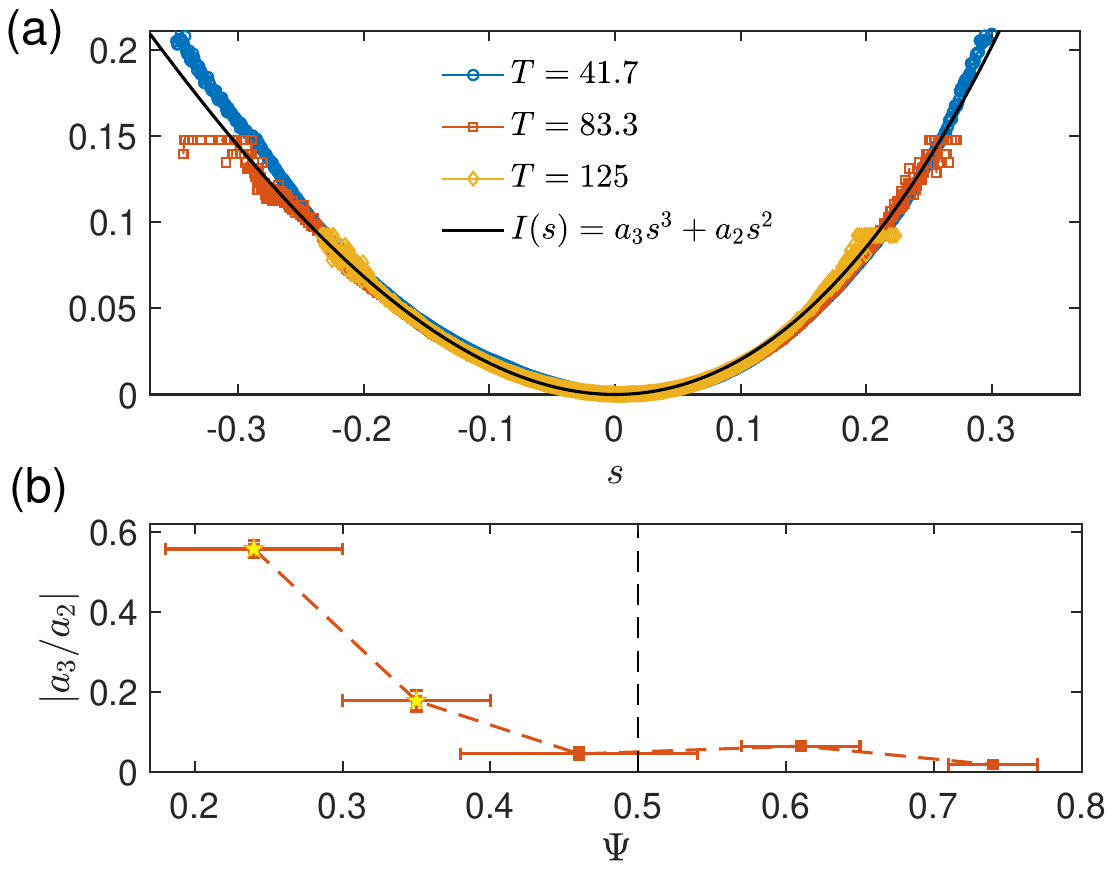}
    \caption{\textbf{Rate function:}
      \textit{Top:} Plot of the rate function $I(s)$, defined in \eq{eq:LDP},
      versus $s$ for $\RC$.
      The plot is obtained by plotting the LHS of \eq{eq:LDP}
      for three large values of  $T\approx 42$, $83$ and $125$.
      The data collapses on a $T$-independent curve. 
      The black curve is the least-square fit to the data using
      the polynomial $\atwo s^2 + \ath s^3$.
      \textit{Bottom:} The ratio
      $\beta = \mid \ath/\atwo \mid$
      as a function of $\Psi$.
      The vertical error bars are derived from the $95$\% confidence intervals
      of the fitted values of $\ath$ and $\atwo$ while the horizontal
      errorbars represent the standard error in $\Psi$.
      }
    \label{fig:LDP}
\end{figure}

From our DNS we calculate the PDF of $R_x/T$ for several
large values of $T$ and find that  
the left hand side of \eq{eq:LDP} converges to a
$T$-independent  function $I(s)$.
For \RC, this is shown in \subfig{fig:LDP}{a}; 
see the \Fig{fig:LDPrest} for similar plots for the other runs. 
We find that, for \RC, $I(s)$ can be well
approximated, in the least-square sense, by truncating the power series
expansion at $s^3$, i.e.,
$I(s) \approx \atwo s^2 + \ath s^3 $~\footnote{The rate function $I(s)$
must be convex up, hence in principle there must be at least a $s^4$ term but
to the range of $s$ over which we have collapsed data
this term is negligible.} -- shown by the dashed curve in \subfig{fig:LDP}{a} --
and thus,
\begin{equation}
  P\left(\frac{R_x}{T} \right) \asymp \exp\left[
    -\atwo\frac{R^2_x}{T} - \ath\frac{R_x^3}{T^2}\right]\/.
  \label{eq:Pasymp}
\end{equation}
Neither is the PDF $P(R_x/T)$ a Gaussian nor is 
there a unique dynamic scaling exponent.
Considering only the second term within the square brackets
in \eq{eq:Pasymp} we obtain
$z = 3/2$ and $\avg{R_x^2} \sim t^{4/3}$. 
Hence superdiffusion, with exponent $\gamma = 4/3$ in \eq{eq:Taylor},
follows from the cubic term in the rate function.
Furthermore, the superdiffusive scaling holds as long as the second
term within the square bracket in  \eq{eq:Pasymp} dominates over the
first, i.e.,  up to time  $T < \Tast$, where
$\Tast \equiv R_x\mid\ath/\atwo\mid$. 
From our DNS we plot the ratio
$\beta \equiv \mid \ath/\atwo \mid$ as a function
of the solenoidal ratio $\Psi$ in \subfig{fig:LDP}{b}.
For $\Psi \gtrsim 1/2$, $\beta$  is close to zero,
the rate function is well approximated by a quadratic and $P(R_x/T)$
is well approximated by a Gaussian.
These are the runs dominated by the solenoidal component of the velocity
and shows Taylor diffusion. 
%---------------------------------------------------
%---------------------------------------------------
\begin{figure*}
    \centering
    \includegraphics[width=0.99\textwidth]{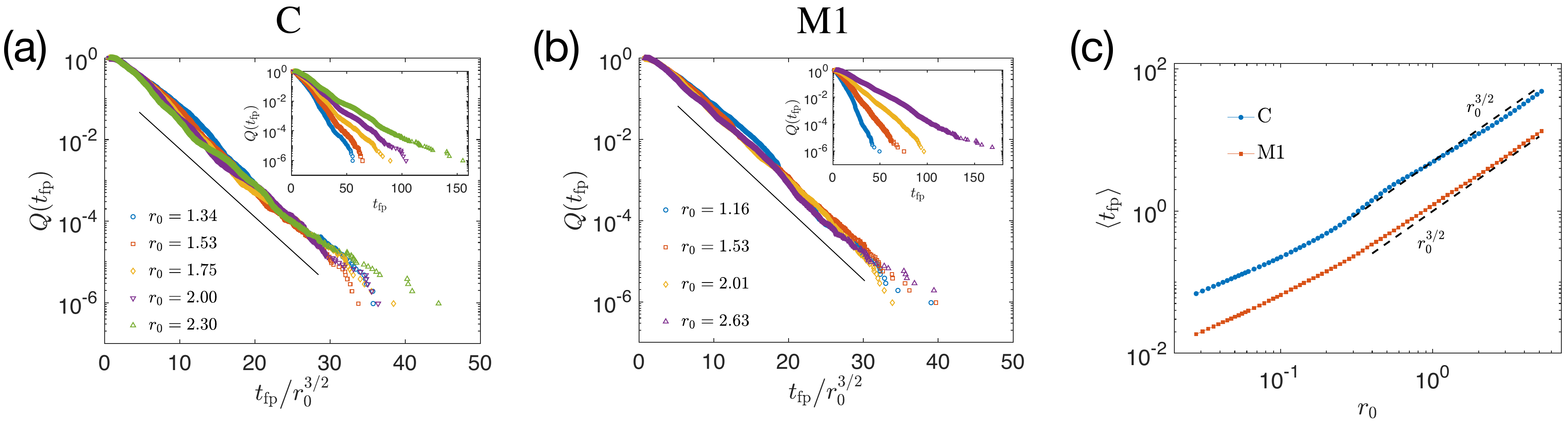}
    \caption{\textbf{First-passage times:}
      Complementary cumulative PDF  of first-passage times, $Q(S)$, for
      different target radii, $\rz>1$, within the superdiffusive regime
      for the runs \RC and \RMo in \textbf{a} and \textbf{b} respectively.  
      The insets show the CCPDFs and the main plots show their
      collapse upon rescaling $S$ by $\rz^{3/2}$.  
      \textbf{(c)} Log-log plots of $\avg{\tfp}$ versus $\rz$ in the
      \RC and \RMo. 
      The dashed lines have slopes equal to $3/2$.
        \label{fig:FPT} }

\end{figure*}
%---------------------------------------------------

It is well known that different mechanism of
anomalous diffusion can give the same exponent.
Hence, to uncover the mechanism of superdiffusion we look at the
scaling of the PDF, $p(\tfp; \rz)$, of the \textit{first-passage times},
$\tfp$, defined as the time taken for the (magnitude of the)
displacement, $R$,
of a tracer to exceed a fixed threshold $\rz$,
\textit{for the first time}~\cite{Redner2001guide}.
To circumvent binning errors we use the rank-order method
to calculate the complementary
cumulative probability distribution function (CCPDFs),
$Q(\tfp; \rz) \equiv 1-\int_0^{\tfp} p(t; \rz)dt$.
We plot the results for \RC and \RMo, for several values of $\rz$ in
superdiffusive range, in \subfig{fig:FPT}{a}. 
All the CCPDFs have exponential tails, as shown in the insets of
\subfig{fig:FPT}{a} and \subfig{fig:FPT}{b}.
Upon rescaling $\tfp$ by $\rz^{3/2}$ they collapse onto a single universal
curve:
\begin{equation}
  Q(\tfp;\rz) \sim \exp\left(-\frac{\tfp}{\rz^{\alpha}} \right)
  \quad\text{as}\quad
   \tfp \to \infty \/,
  \label{eq:Qexp}
\end{equation}
where $\alpha=3/2$.
Consequently, the mean first passage time
$\avg{\tfp(\rz)} \propto \rz^{\alpha}$.
We confirm~\footnote{The scaling range in \subfig{fig:FPT}{c} is
not large enough to do a proper error analysis. A least-square
fit to the data over a scaling range estimated by the eye
gives an exponent of approximately $1.4$. }
this result for both \RC and \RMo,
in \subfig{fig:FPT}{c}.
Equation \ref{eq:Qexp} is the canonical form of the scaling of the CCPDF of
first-passage times of random-walkers undergoing a
\textit{heterogeneous diffusion process (HDP)}~\cite{Chertsvy2013HetDiff1,
  Cherstvy2013HetDiff2,Xu2020HetDiff, Metzler2014AnoDiff}
with $\alpha=2/\gamma$.
The values of $\alpha$ calculated from $\gamma$
($2/\gamma = 3/2$) using this relation
agrees $\alpha=3/2$ obtained from the data collapse of the
PDF of first passage time. 
This affirms  that the superdiffusive transport of
tracers in \RC and
\RMo can be modelled by a heterogeneous diffusion process. 
By contrast in \RS, which does not show superdiffusion,
we obtain  $\avg{\tfp}\sim \rz^2$ for large  $\rz$,
implying simple diffusion.

Several comments are now in order.

We observe superdiffusion over a time window
that increases as solenoidality decreases, or the
contributions of the shocks to the flow increases,
but at late times we always recover simple diffusion.
Therefore, we conjecture that for fully irrotational flows
we would observe:
$\avg{R^2(t)} \sim t^{4/3}$,  
even at late times -- simple diffusion is never observed.
We confirm this by doing simulations of the Burgers
equation in 2D -- see \ref{app:Burgers}. 

We demonstrate that the PDF of displacement of tracers, in all cases,
obeys the large deviation principle.
For large solenoidality the rate function is well-approximated by a
parabola.
As solenoidality decreases the cubic term becomes non-negligible
-- the quantity $\beta$ increases, see \subfig{fig:LDP}{b}.
This is what gives rise to superdiffusion.
Is there are a critical value of $\Psi = \Psic$ such
that $\beta = 0$ for $\Psi > \Psic$ ?
Present simulations do not give a decisive answer. 
Note also that, tracer particles in active turbulence are also shown
to undergo superdiffusion~\cite{Samriddhi2022PRF} with the same
exponent of $4/3$.
It is plausible that the same large-deviation-based argument
may apply there, too. 

We find that heterogeneous diffusion is a reasonable
model for the superdiffusion based on statistics of the first
passage time. But this is not consistent with the fact that
we are solving homogeneous and isotropic turbulence.
We conjecture that presence of large-scale shocks can
make our flow inhomogeneous over the timescale of
the lifetime of the shocks.
At times much longer than the lifetime of all shocks
homogeneity is restored and superdiffusion gives
way to simple diffusion.

Taylor’s result has shaped our understanding of
particle transport in nearly incompressible astrophysical and
geophysical flows.
Our result should prove similarly important for strongly irrotational
astrophysical flows, such as those in molecular
clouds~\cite{Mordecai2004Review, Federrath2012SFR, Orkisz2017SF}.
Superdiffusion spreads particles faster than ordinary diffusion.
In cosmological simulations~\cite{Volker2002Cosmo, Weinberger2022Cosmo}
particle transport is considered as sub-grid-scale process and
usually modelled by a Laplacian operator.
Superdiffusion cannot be captured simply by increasing the diffusion
coefficient; it would be necessary to use fractional differential operator
of order $3/2$.
However, as our simulations are in 2D we cannot yet
claim that particle transport in molecular clouds is likely to
be superdiffusive.
Our results instead provide strong motivation to study  particle
transport in three-dimensional, compressible turbulence dominated by shocks.

In summary, we show that the transport of tracers in compressible
shock-dominated turbulence is superdiffusive over a time window
which increases as the flow becomes less solenoidal
and, for small enough solenoidality, can
exceed the large-eddy turnover time.
At very late times simple diffusion is observed.
We also rationalize this result through a large-deviation
approach and show its possible connection to heterogeneous
diffusion. 

%-----------------------------------
\paragraph*{Acknowledgements:}
  SD thanks the NORDITA visiting PhD Fellowship program for partially
  supporting his visit to NORDITA. 
  SD also acknowledges support from the Department of Physics at the
  Indian Institute of Science, Bangalore, where a major portion of the work was
  carried out.
  RP thanks the ANRF (India) for support.
  DM acknowledges the support of the Swedish Research Council
  Grant No. 638-2013-9243 and 2016-05225.
%\bibliography{../turb_ref.bib}

\begin{thebibliography}{48}%
\makeatletter
\providecommand \@ifxundefined [1]{%
 \@ifx{#1\undefined}
}%
\providecommand \@ifnum [1]{%
 \ifnum #1\expandafter \@firstoftwo
 \else \expandafter \@secondoftwo
 \fi
}%
\providecommand \@ifx [1]{%
 \ifx #1\expandafter \@firstoftwo
 \else \expandafter \@secondoftwo
 \fi
}%
\providecommand \natexlab [1]{#1}%
\providecommand \enquote  [1]{``#1''}%
\providecommand \bibnamefont  [1]{#1}%
\providecommand \bibfnamefont [1]{#1}%
\providecommand \citenamefont [1]{#1}%
\providecommand \href@noop [0]{\@secondoftwo}%
\providecommand \href [0]{\begingroup \@sanitize@url \@href}%
\providecommand \@href[1]{\@@startlink{#1}\@@href}%
\providecommand \@@href[1]{\endgroup#1\@@endlink}%
\providecommand \@sanitize@url [0]{\catcode `\\12\catcode `\$12\catcode
  `\&12\catcode `\#12\catcode `\^12\catcode `\_12\catcode `\%12\relax}%
\providecommand \@@startlink[1]{}%
\providecommand \@@endlink[0]{}%
\providecommand \url  [0]{\begingroup\@sanitize@url \@url }%
\providecommand \@url [1]{\endgroup\@href {#1}{\urlprefix }}%
\providecommand \urlprefix  [0]{URL }%
\providecommand \Eprint [0]{\href }%
\providecommand \doibase [0]{http://dx.doi.org/}%
\providecommand \selectlanguage [0]{\@gobble}%
\providecommand \bibinfo  [0]{\@secondoftwo}%
\providecommand \bibfield  [0]{\@secondoftwo}%
\providecommand \translation [1]{[#1]}%
\providecommand \BibitemOpen [0]{}%
\providecommand \bibitemStop [0]{}%
\providecommand \bibitemNoStop [0]{.\EOS\space}%
\providecommand \EOS [0]{\spacefactor3000\relax}%
\providecommand \BibitemShut  [1]{\csname bibitem#1\endcsname}%
\let\auto@bib@innerbib\@empty
%</preamble>
\bibitem [{\citenamefont {Kolmogorov}(1941)}]{K41}%
  \BibitemOpen
  \bibfield  {author} {\bibinfo {author} {\bibfnamefont {A.~N.}\ \bibnamefont
  {Kolmogorov}},\ }\bibfield  {title} {\enquote {\bibinfo {title} {On the
  degeneration of isotropic turbulence in an incompressible viscous fluids},}\
  }\href@noop {} {\bibfield  {journal} {\bibinfo  {journal} {Dokl. Akad. Nauk.
  SSSR}\ }\textbf {\bibinfo {volume} {31}},\ \bibinfo {pages} {5385--541}
  (\bibinfo {year} {1941})}\BibitemShut {NoStop}%
\bibitem [{\citenamefont {Frisch}(1996)}]{frisch}%
  \BibitemOpen
  \bibfield  {author} {\bibinfo {author} {\bibfnamefont {U.}~\bibnamefont
  {Frisch}},\ }\href@noop {} {\emph {\bibinfo {title} {Turbulence: The Legacy
  of A.N. Kolmogorov}}}\ (\bibinfo  {publisher} {Cambridge University Press,
  Cambridge, England},\ \bibinfo {year} {1996})\BibitemShut {NoStop}%
\bibitem [{\citenamefont {Richardson}(1926)}]{richardson1926atmospheric}%
  \BibitemOpen
  \bibfield  {author} {\bibinfo {author} {\bibfnamefont {Lewis~Fry}\
  \bibnamefont {Richardson}},\ }\bibfield  {title} {\enquote {\bibinfo {title}
  {Atmospheric diffusion shown on a distance-neighbour graph},}\ }\href@noop {}
  {\bibfield  {journal} {\bibinfo  {journal} {Proceedings of the Royal Society
  of London. Series A, Containing Papers of a Mathematical and Physical
  Character}\ }\textbf {\bibinfo {volume} {110}},\ \bibinfo {pages} {709--737}
  (\bibinfo {year} {1926})}\BibitemShut {NoStop}%
\bibitem [{\citenamefont {Taylor}(1922)}]{Taylor1922Diffusion}%
  \BibitemOpen
  \bibfield  {author} {\bibinfo {author} {\bibfnamefont {Geoffrey~I}\
  \bibnamefont {Taylor}},\ }\bibfield  {title} {\enquote {\bibinfo {title}
  {Diffusion by continuous movements},}\ }\href@noop {} {\bibfield  {journal}
  {\bibinfo  {journal} {Proceedings of the london mathematical society}\
  }\textbf {\bibinfo {volume} {2}},\ \bibinfo {pages} {196--212} (\bibinfo
  {year} {1922})}\BibitemShut {NoStop}%
\bibitem [{\citenamefont {Mac~Low}\ and\ \citenamefont
  {Klessen}(2004)}]{Mordecai2004Review}%
  \BibitemOpen
  \bibfield  {author} {\bibinfo {author} {\bibfnamefont {Mordecai-Mark}\
  \bibnamefont {Mac~Low}}\ and\ \bibinfo {author} {\bibfnamefont {Ralf~S.}\
  \bibnamefont {Klessen}},\ }\bibfield  {title} {\enquote {\bibinfo {title}
  {Control of star formation by supersonic turbulence},}\ }\href {\doibase
  10.1103/RevModPhys.76.125} {\bibfield  {journal} {\bibinfo  {journal} {Rev.
  Mod. Phys.}\ }\textbf {\bibinfo {volume} {76}},\ \bibinfo {pages} {125--194}
  (\bibinfo {year} {2004})}\BibitemShut {NoStop}%
\bibitem [{\citenamefont {Elmegreen}\ and\ \citenamefont
  {Scalo}(2004)}]{Elmegreen2004Review_ISM}%
  \BibitemOpen
  \bibfield  {author} {\bibinfo {author} {\bibfnamefont {Bruce~G.}\
  \bibnamefont {Elmegreen}}\ and\ \bibinfo {author} {\bibfnamefont {John}\
  \bibnamefont {Scalo}},\ }\bibfield  {title} {\enquote {\bibinfo {title}
  {{Interstellar Turbulence I: Observations and Processes}},}\ }\href {\doibase
  https://doi.org/10.1146/annurev.astro.41.011802.094859} {\bibfield  {journal}
  {\bibinfo  {journal} {Annual Review of Astronomy and Astrophysics}\ }\textbf
  {\bibinfo {volume} {42}},\ \bibinfo {pages} {211--273} (\bibinfo {year}
  {2004})}\BibitemShut {NoStop}%
\bibitem [{\citenamefont {Larson}(1981)}]{Larson1981SF}%
  \BibitemOpen
  \bibfield  {author} {\bibinfo {author} {\bibfnamefont {R.~B.}\ \bibnamefont
  {Larson}},\ }\bibfield  {title} {\enquote {\bibinfo {title} {Turbulence and
  star formation in molecular clouds},}\ }\href
  {https://articles.adsabs.harvard.edu/full/1981MNRAS.194..809L} {\bibfield
  {journal} {\bibinfo  {journal} {Mon. Not. R. Astron. Soc.}\ }\textbf
  {\bibinfo {volume} {194}},\ \bibinfo {pages} {809--826} (\bibinfo {year}
  {1981})}\BibitemShut {NoStop}%
\bibitem [{\citenamefont {Padoan}\ and\ \citenamefont {Åke
  Nordlund}(2011)}]{Padoan2011SFR}%
  \BibitemOpen
  \bibfield  {author} {\bibinfo {author} {\bibfnamefont {Paolo}\ \bibnamefont
  {Padoan}}\ and\ \bibinfo {author} {\bibnamefont {Åke Nordlund}},\ }\bibfield
   {title} {\enquote {\bibinfo {title} {The star formation rate of supersonic
  magnetohydrodynamic turbulence},}\ }\href {\doibase
  10.1088/0004-637X/730/1/40} {\bibfield  {journal} {\bibinfo  {journal} {The
  Astrophysical Journal}\ }\textbf {\bibinfo {volume} {730}},\ \bibinfo {pages}
  {40} (\bibinfo {year} {2011})}\BibitemShut {NoStop}%
\bibitem [{\citenamefont {Federrath}\ and\ \citenamefont
  {Klessen}(2012)}]{Federrath2012SFR}%
  \BibitemOpen
  \bibfield  {author} {\bibinfo {author} {\bibfnamefont {Christoph}\
  \bibnamefont {Federrath}}\ and\ \bibinfo {author} {\bibfnamefont {Ralf~S.}\
  \bibnamefont {Klessen}},\ }\bibfield  {title} {\enquote {\bibinfo {title}
  {The star formation rate of turbulent magnetized clouds: Comparing theory,
  simulations, and observations},}\ }\href {\doibase
  10.1088/0004-637X/761/2/156} {\bibfield  {journal} {\bibinfo  {journal} {The
  Astrophysical Journal}\ }\textbf {\bibinfo {volume} {761}},\ \bibinfo {pages}
  {156} (\bibinfo {year} {2012})}\BibitemShut {NoStop}%
\bibitem [{\citenamefont {{Orkisz, Jan H.}}\ \emph {et~al.}(2017)\citenamefont
  {{Orkisz, Jan H.}}, \citenamefont {{Pety, Jérôme}}, \citenamefont {{Gerin,
  Maryvonne}}, \citenamefont {{Bron, Emeric}}, \citenamefont {{Guzmán, Viviana
  V.}}, \citenamefont {{Bardeau, Sébastien}}, \citenamefont {{Goicoechea,
  Javier R.}}, \citenamefont {{Gratier, Pierre}}, \citenamefont {{Le Petit,
  Franck}}, \citenamefont {{Levrier, François}}, \citenamefont {{Liszt,
  Harvey}}, \citenamefont {{Öberg, Karin}}, \citenamefont {{Peretto,
  Nicolas}}, \citenamefont {{Roueff, Evelyne}}, \citenamefont {{Sievers,
  Albrecht}},\ and\ \citenamefont {{Tremblin, Pascal}}}]{Orkisz2017SF}%
  \BibitemOpen
  \bibfield  {author} {\bibinfo {author} {\bibnamefont {{Orkisz, Jan H.}}},
  \bibinfo {author} {\bibnamefont {{Pety, Jérôme}}}, \bibinfo {author}
  {\bibnamefont {{Gerin, Maryvonne}}}, \bibinfo {author} {\bibnamefont {{Bron,
  Emeric}}}, \bibinfo {author} {\bibnamefont {{Guzmán, Viviana V.}}}, \bibinfo
  {author} {\bibnamefont {{Bardeau, Sébastien}}}, \bibinfo {author}
  {\bibnamefont {{Goicoechea, Javier R.}}}, \bibinfo {author} {\bibnamefont
  {{Gratier, Pierre}}}, \bibinfo {author} {\bibnamefont {{Le Petit, Franck}}},
  \bibinfo {author} {\bibnamefont {{Levrier, François}}}, \bibinfo {author}
  {\bibnamefont {{Liszt, Harvey}}}, \bibinfo {author} {\bibnamefont {{Öberg,
  Karin}}}, \bibinfo {author} {\bibnamefont {{Peretto, Nicolas}}}, \bibinfo
  {author} {\bibnamefont {{Roueff, Evelyne}}}, \bibinfo {author} {\bibnamefont
  {{Sievers, Albrecht}}}, \ and\ \bibinfo {author} {\bibnamefont {{Tremblin,
  Pascal}}},\ }\bibfield  {title} {\enquote {\bibinfo {title} {Turbulence and
  star formation efficiency in molecular clouds: solenoidal versus compressive
  motions in orion b},}\ }\href {\doibase 10.1051/0004-6361/201629220}
  {\bibfield  {journal} {\bibinfo  {journal} {A\&A}\ }\textbf {\bibinfo
  {volume} {599}},\ \bibinfo {pages} {A99} (\bibinfo {year}
  {2017})}\BibitemShut {NoStop}%
\bibitem [{\citenamefont {Pan}\ \emph {et~al.}(2011)\citenamefont {Pan},
  \citenamefont {Padoan}, \citenamefont {Scalo}, \citenamefont {Kritsuk},\ and\
  \citenamefont {Norman}}]{Pan2011Planetesimal_Clust}%
  \BibitemOpen
  \bibfield  {author} {\bibinfo {author} {\bibfnamefont {Liubin}\ \bibnamefont
  {Pan}}, \bibinfo {author} {\bibfnamefont {Paolo}\ \bibnamefont {Padoan}},
  \bibinfo {author} {\bibfnamefont {John}\ \bibnamefont {Scalo}}, \bibinfo
  {author} {\bibfnamefont {Alexei~G.}\ \bibnamefont {Kritsuk}}, \ and\ \bibinfo
  {author} {\bibfnamefont {Michael~L.}\ \bibnamefont {Norman}},\ }\bibfield
  {title} {\enquote {\bibinfo {title} {Turbulent clustering of protoplanetary
  dust and planetesimal formation},}\ }\href {\doibase
  10.1088/0004-637X/740/1/6} {\bibfield  {journal} {\bibinfo  {journal} {The
  Astrophysical Journal}\ }\textbf {\bibinfo {volume} {740}},\ \bibinfo {pages}
  {6} (\bibinfo {year} {2011})}\BibitemShut {NoStop}%
\bibitem [{\citenamefont {Boldyrev}\ \emph
  {et~al.}(2002{\natexlab{a}})\citenamefont {Boldyrev}, \citenamefont {Åke
  Nordlund},\ and\ \citenamefont {Padoan}}]{Boldyrev2002SF_MC}%
  \BibitemOpen
  \bibfield  {author} {\bibinfo {author} {\bibfnamefont {Stanislav}\
  \bibnamefont {Boldyrev}}, \bibinfo {author} {\bibnamefont {Åke Nordlund}}, \
  and\ \bibinfo {author} {\bibfnamefont {Paolo}\ \bibnamefont {Padoan}},\
  }\bibfield  {title} {\enquote {\bibinfo {title} {Scaling relations of
  supersonic turbulence in star-forming molecular clouds},}\ }\href {\doibase
  10.1086/340758} {\bibfield  {journal} {\bibinfo  {journal} {The Astrophysical
  Journal}\ }\textbf {\bibinfo {volume} {573}},\ \bibinfo {pages} {678}
  (\bibinfo {year} {2002}{\natexlab{a}})}\BibitemShut {NoStop}%
\bibitem [{\citenamefont {Boldyrev}\ \emph
  {et~al.}(2002{\natexlab{b}})\citenamefont {Boldyrev}, \citenamefont
  {Nordlund},\ and\ \citenamefont {Padoan}}]{Boldyrev2002SupersonicISM}%
  \BibitemOpen
  \bibfield  {author} {\bibinfo {author} {\bibfnamefont {Stanislav}\
  \bibnamefont {Boldyrev}}, \bibinfo {author} {\bibfnamefont {\AA{}ke}\
  \bibnamefont {Nordlund}}, \ and\ \bibinfo {author} {\bibfnamefont {Paolo}\
  \bibnamefont {Padoan}},\ }\bibfield  {title} {\enquote {\bibinfo {title}
  {Supersonic turbulence and structure of interstellar molecular clouds},}\
  }\href {\doibase 10.1103/PhysRevLett.89.031102} {\bibfield  {journal}
  {\bibinfo  {journal} {Phys. Rev. Lett.}\ }\textbf {\bibinfo {volume} {89}},\
  \bibinfo {pages} {031102} (\bibinfo {year} {2002}{\natexlab{b}})}\BibitemShut
  {NoStop}%
\bibitem [{\citenamefont {Kritsuk}\ \emph {et~al.}(2007)\citenamefont
  {Kritsuk}, \citenamefont {Norman}, \citenamefont {Padoan},\ and\
  \citenamefont {Wagner}}]{Kritsuk2007ISM}%
  \BibitemOpen
  \bibfield  {author} {\bibinfo {author} {\bibfnamefont {Alexei~G.}\
  \bibnamefont {Kritsuk}}, \bibinfo {author} {\bibfnamefont {Michael~L.}\
  \bibnamefont {Norman}}, \bibinfo {author} {\bibfnamefont {Paolo}\
  \bibnamefont {Padoan}}, \ and\ \bibinfo {author} {\bibfnamefont {Rick}\
  \bibnamefont {Wagner}},\ }\bibfield  {title} {\enquote {\bibinfo {title} {The
  statistics of supersonic isothermal turbulence},}\ }\href {\doibase
  10.1086/519443} {\bibfield  {journal} {\bibinfo  {journal} {The Astrophysical
  Journal}\ }\textbf {\bibinfo {volume} {665}},\ \bibinfo {pages} {416}
  (\bibinfo {year} {2007})}\BibitemShut {NoStop}%
\bibitem [{\citenamefont {Federrath}\ \emph {et~al.}(2009)\citenamefont
  {Federrath}, \citenamefont {Klessen},\ and\ \citenamefont
  {Schmidt}}]{Federrath2009FractalDensity}%
  \BibitemOpen
  \bibfield  {author} {\bibinfo {author} {\bibfnamefont {Christoph}\
  \bibnamefont {Federrath}}, \bibinfo {author} {\bibfnamefont {Ralf~S.}\
  \bibnamefont {Klessen}}, \ and\ \bibinfo {author} {\bibfnamefont {Wolfram}\
  \bibnamefont {Schmidt}},\ }\bibfield  {title} {\enquote {\bibinfo {title}
  {The fractal density structure in supersonic isothermal turbulence:
  Solenoidal versus compressive energy injection},}\ }\href {\doibase
  10.1088/0004-637X/692/1/364} {\bibfield  {journal} {\bibinfo  {journal} {The
  Astrophysical Journal}\ }\textbf {\bibinfo {volume} {692}},\ \bibinfo {pages}
  {364} (\bibinfo {year} {2009})}\BibitemShut {NoStop}%
\bibitem [{\citenamefont {Federrath}\ \emph {et~al.}(2008)\citenamefont
  {Federrath}, \citenamefont {Klessen},\ and\ \citenamefont
  {Schmidt}}]{Federrath2008DensityPDF}%
  \BibitemOpen
  \bibfield  {author} {\bibinfo {author} {\bibfnamefont {Christoph}\
  \bibnamefont {Federrath}}, \bibinfo {author} {\bibfnamefont {Ralf~S.}\
  \bibnamefont {Klessen}}, \ and\ \bibinfo {author} {\bibfnamefont {Wolfram}\
  \bibnamefont {Schmidt}},\ }\bibfield  {title} {\enquote {\bibinfo {title}
  {The density probability distribution in compressible isothermal turbulence:
  Solenoidal versus compressive forcing},}\ }\href {\doibase 10.1086/595280}
  {\bibfield  {journal} {\bibinfo  {journal} {The Astrophysical Journal}\
  }\textbf {\bibinfo {volume} {688}},\ \bibinfo {pages} {L79} (\bibinfo {year}
  {2008})}\BibitemShut {NoStop}%
\bibitem [{\citenamefont {Federrath}\ \emph {et~al.}(2011)\citenamefont
  {Federrath}, \citenamefont {Chabrier}, \citenamefont {Schober}, \citenamefont
  {Banerjee}, \citenamefont {Klessen},\ and\ \citenamefont
  {Schleicher}}]{Federrath2011Machno}%
  \BibitemOpen
  \bibfield  {author} {\bibinfo {author} {\bibfnamefont {C.}~\bibnamefont
  {Federrath}}, \bibinfo {author} {\bibfnamefont {G.}~\bibnamefont {Chabrier}},
  \bibinfo {author} {\bibfnamefont {J.}~\bibnamefont {Schober}}, \bibinfo
  {author} {\bibfnamefont {R.}~\bibnamefont {Banerjee}}, \bibinfo {author}
  {\bibfnamefont {R.~S.}\ \bibnamefont {Klessen}}, \ and\ \bibinfo {author}
  {\bibfnamefont {D.~R.~G.}\ \bibnamefont {Schleicher}},\ }\bibfield  {title}
  {\enquote {\bibinfo {title} {Mach number dependence of turbulent magnetic
  field amplification: Solenoidal versus compressive flows},}\ }\href {\doibase
  10.1103/PhysRevLett.107.114504} {\bibfield  {journal} {\bibinfo  {journal}
  {Phys. Rev. Lett.}\ }\textbf {\bibinfo {volume} {107}},\ \bibinfo {pages}
  {114504} (\bibinfo {year} {2011})}\BibitemShut {NoStop}%
\bibitem [{\citenamefont {Federrath}(2013)}]{Federrath2013Universality}%
  \BibitemOpen
  \bibfield  {author} {\bibinfo {author} {\bibfnamefont {Christoph}\
  \bibnamefont {Federrath}},\ }\bibfield  {title} {\enquote {\bibinfo {title}
  {{On the universality of supersonic turbulence}},}\ }\href {\doibase
  10.1093/mnras/stt1644} {\bibfield  {journal} {\bibinfo  {journal} {Monthly
  Notices of the Royal Astronomical Society}\ }\textbf {\bibinfo {volume}
  {436}},\ \bibinfo {pages} {1245--1257} (\bibinfo {year} {2013})},\ \Eprint
  {http://arxiv.org/abs/https://academic.oup.com/mnras/article-pdf/436/2/1245/13763684/stt1644.pdf}
  {https://academic.oup.com/mnras/article-pdf/436/2/1245/13763684/stt1644.pdf}
  \BibitemShut {NoStop}%
\bibitem [{\citenamefont {Schmidt}\ \emph {et~al.}(2008)\citenamefont
  {Schmidt}, \citenamefont {Federrath},\ and\ \citenamefont
  {Klessen}}]{Schmidt2008Universal}%
  \BibitemOpen
  \bibfield  {author} {\bibinfo {author} {\bibfnamefont {Wolfram}\ \bibnamefont
  {Schmidt}}, \bibinfo {author} {\bibfnamefont {Christoph}\ \bibnamefont
  {Federrath}}, \ and\ \bibinfo {author} {\bibfnamefont {Ralf}\ \bibnamefont
  {Klessen}},\ }\bibfield  {title} {\enquote {\bibinfo {title} {Is the scaling
  of supersonic turbulence universal?}}\ }\href {\doibase
  10.1103/PhysRevLett.101.194505} {\bibfield  {journal} {\bibinfo  {journal}
  {Phys. Rev. Lett.}\ }\textbf {\bibinfo {volume} {101}},\ \bibinfo {pages}
  {194505} (\bibinfo {year} {2008})}\BibitemShut {NoStop}%
\bibitem [{\citenamefont {Klessen}\ and\ \citenamefont
  {Lin}(2003)}]{Klessen2003Diffusion}%
  \BibitemOpen
  \bibfield  {author} {\bibinfo {author} {\bibfnamefont {Ralf~S.}\ \bibnamefont
  {Klessen}}\ and\ \bibinfo {author} {\bibfnamefont {Douglas N.~C.}\
  \bibnamefont {Lin}},\ }\bibfield  {title} {\enquote {\bibinfo {title}
  {Diffusion in supersonic turbulent compressible flows},}\ }\href {\doibase
  10.1103/PhysRevE.67.046311} {\bibfield  {journal} {\bibinfo  {journal} {Phys.
  Rev. E}\ }\textbf {\bibinfo {volume} {67}},\ \bibinfo {pages} {046311}
  (\bibinfo {year} {2003})}\BibitemShut {NoStop}%
\bibitem [{\citenamefont {Zhang}\ \emph {et~al.}(2016)\citenamefont {Zhang},
  \citenamefont {Liu}, \citenamefont {Ma},\ and\ \citenamefont
  {Xiao}}]{Zhang2016Inertial}%
  \BibitemOpen
  \bibfield  {author} {\bibinfo {author} {\bibfnamefont {Qingqing}\
  \bibnamefont {Zhang}}, \bibinfo {author} {\bibfnamefont {Han}\ \bibnamefont
  {Liu}}, \bibinfo {author} {\bibfnamefont {Zongqiang}\ \bibnamefont {Ma}}, \
  and\ \bibinfo {author} {\bibfnamefont {Zuoli}\ \bibnamefont {Xiao}},\
  }\bibfield  {title} {\enquote {\bibinfo {title} {{Preferential concentration
  of heavy particles in compressible isotropic turbulence}},}\ }\href {\doibase
  10.1063/1.4948810} {\bibfield  {journal} {\bibinfo  {journal} {Physics of
  Fluids}\ }\textbf {\bibinfo {volume} {28}},\ \bibinfo {pages} {055104}
  (\bibinfo {year} {2016})}\BibitemShut {NoStop}%
\bibitem [{\citenamefont {Zhang}\ and\ \citenamefont
  {Xiao}(2018)}]{Zhang2018SinglePl}%
  \BibitemOpen
  \bibfield  {author} {\bibinfo {author} {\bibfnamefont {Qingqing}\
  \bibnamefont {Zhang}}\ and\ \bibinfo {author} {\bibfnamefont {Zuoli}\
  \bibnamefont {Xiao}},\ }\bibfield  {title} {\enquote {\bibinfo {title}
  {{Single-particle dispersion in compressible turbulence}},}\ }\href {\doibase
  10.1063/1.5011083} {\bibfield  {journal} {\bibinfo  {journal} {Physics of
  Fluids}\ }\textbf {\bibinfo {volume} {30}},\ \bibinfo {pages} {040904}
  (\bibinfo {year} {2018})},\ \Eprint
  {http://arxiv.org/abs/https://pubs.aip.org/aip/pof/article-pdf/doi/10.1063/1.5011083/16673987/040904\_1\_online.pdf}
  {https://pubs.aip.org/aip/pof/article-pdf/doi/10.1063/1.5011083/16673987/040904\_1\_online.pdf}
  \BibitemShut {NoStop}%
\bibitem [{\citenamefont {De}\ \emph {et~al.}(2024)\citenamefont {De},
  \citenamefont {Mitra},\ and\ \citenamefont {Pandit}}]{de2024uncovering}%
  \BibitemOpen
  \bibfield  {author} {\bibinfo {author} {\bibfnamefont {Sadhitro}\
  \bibnamefont {De}}, \bibinfo {author} {\bibfnamefont {Dhrubaditya}\
  \bibnamefont {Mitra}}, \ and\ \bibinfo {author} {\bibfnamefont {Rahul}\
  \bibnamefont {Pandit}},\ }\bibfield  {title} {\enquote {\bibinfo {title}
  {Uncovering the multifractality of lagrangian pair dispersion in
  shock-dominated turbulence},}\ }\href@noop {} {\bibfield  {journal} {\bibinfo
   {journal} {Physical Review Research}\ }\textbf {\bibinfo {volume} {6}},\
  \bibinfo {pages} {L022032} (\bibinfo {year} {2024})}\BibitemShut {NoStop}%
\bibitem [{\citenamefont {De}\ \emph {et~al.}(2026)\citenamefont {De},
  \citenamefont {Mitra},\ and\ \citenamefont {Pandit}}]{de2026intermittency}%
  \BibitemOpen
  \bibfield  {author} {\bibinfo {author} {\bibfnamefont {Sadhitro}\
  \bibnamefont {De}}, \bibinfo {author} {\bibfnamefont {Dhrubaditya}\
  \bibnamefont {Mitra}}, \ and\ \bibinfo {author} {\bibfnamefont {Rahul}\
  \bibnamefont {Pandit}},\ }\bibfield  {title} {\enquote {\bibinfo {title}
  {Intermittency and nonuniversality of pair dispersion in isothermal
  compressible turbulence},}\ }\href {\doibase 10.1103/kr1z-rf6b} {\bibfield
  {journal} {\bibinfo  {journal} {Phys. Rev. Res.}\ }\textbf {\bibinfo {volume}
  {8}},\ \bibinfo {pages} {L012037} (\bibinfo {year} {2026})}\BibitemShut
  {NoStop}%
\bibitem [{\citenamefont {Xia}\ \emph {et~al.}(2013)\citenamefont {Xia},
  \citenamefont {Francois}, \citenamefont {Punzmann},\ and\ \citenamefont
  {Shats}}]{Xia2013Diffusion}%
  \BibitemOpen
  \bibfield  {author} {\bibinfo {author} {\bibfnamefont {H.}~\bibnamefont
  {Xia}}, \bibinfo {author} {\bibfnamefont {N.}~\bibnamefont {Francois}},
  \bibinfo {author} {\bibfnamefont {H.}~\bibnamefont {Punzmann}}, \ and\
  \bibinfo {author} {\bibfnamefont {M.}~\bibnamefont {Shats}},\ }\bibfield
  {title} {\enquote {\bibinfo {title} {Lagrangian scale of particle dispersion
  in turbulence},}\ }\href {\doibase 10.1038/ncomms30131} {\bibfield  {journal}
  {\bibinfo  {journal} {Nat. Commun.}\ }\textbf {\bibinfo {volume} {4}}
  (\bibinfo {year} {2013}),\ 10.1038/ncomms30131}\BibitemShut {NoStop}%
\bibitem [{\citenamefont {Xia}\ \emph {et~al.}(2014)\citenamefont {Xia},
  \citenamefont {Francois}, \citenamefont {Punzmann},\ and\ \citenamefont
  {Shats}}]{Xia2014Taylor}%
  \BibitemOpen
  \bibfield  {author} {\bibinfo {author} {\bibfnamefont {H.}~\bibnamefont
  {Xia}}, \bibinfo {author} {\bibfnamefont {N.}~\bibnamefont {Francois}},
  \bibinfo {author} {\bibfnamefont {H.}~\bibnamefont {Punzmann}}, \ and\
  \bibinfo {author} {\bibfnamefont {M.}~\bibnamefont {Shats}},\ }\bibfield
  {title} {\enquote {\bibinfo {title} {Taylor particle dispersion during
  transition to fully developed two-dimensional turbulence},}\ }\href {\doibase
  10.1103/PhysRevLett.112.104501} {\bibfield  {journal} {\bibinfo  {journal}
  {Phys. Rev. Lett.}\ }\textbf {\bibinfo {volume} {112}},\ \bibinfo {pages}
  {104501} (\bibinfo {year} {2014})}\BibitemShut {NoStop}%
\bibitem [{\citenamefont {Mordant}\ \emph {et~al.}(2004)\citenamefont
  {Mordant}, \citenamefont {Lévêque},\ and\ \citenamefont
  {Pinton}}]{Mordant2004Diffusion}%
  \BibitemOpen
  \bibfield  {author} {\bibinfo {author} {\bibfnamefont {Nicolas}\ \bibnamefont
  {Mordant}}, \bibinfo {author} {\bibfnamefont {Emmanuel}\ \bibnamefont
  {Lévêque}}, \ and\ \bibinfo {author} {\bibfnamefont {Jean-François}\
  \bibnamefont {Pinton}},\ }\bibfield  {title} {\enquote {\bibinfo {title}
  {{Experimental and numerical study of the Lagrangian dynamics of high
  Reynolds turbulence}},}\ }\href {\doibase 10.1088/1367-2630/6/1/116}
  {\bibfield  {journal} {\bibinfo  {journal} {New Journal of Physics}\ }\textbf
  {\bibinfo {volume} {6}},\ \bibinfo {pages} {116} (\bibinfo {year}
  {2004})}\BibitemShut {NoStop}%
\bibitem [{\citenamefont {Müller}\ and\ \citenamefont
  {Busse}(2007)}]{Muller2007DiffusionMHD}%
  \BibitemOpen
  \bibfield  {author} {\bibinfo {author} {\bibfnamefont {W.-C.}\ \bibnamefont
  {Müller}}\ and\ \bibinfo {author} {\bibfnamefont {A.}~\bibnamefont
  {Busse}},\ }\bibfield  {title} {\enquote {\bibinfo {title} {Diffusion and
  dispersion of passive tracers: Navier-stokes vs. mhd turbulence},}\ }\href
  {\doibase 10.1209/0295-5075/78/14003} {\bibfield  {journal} {\bibinfo
  {journal} {Europhysics Letters}\ }\textbf {\bibinfo {volume} {78}},\ \bibinfo
  {pages} {14003} (\bibinfo {year} {2007})}\BibitemShut {NoStop}%
\bibitem [{\citenamefont {Busse}\ \emph {et~al.}(2007)\citenamefont {Busse},
  \citenamefont {Müller}, \citenamefont {Homann},\ and\ \citenamefont
  {Grauer}}]{Busse2007DiffusionMHD}%
  \BibitemOpen
  \bibfield  {author} {\bibinfo {author} {\bibfnamefont {Angela}\ \bibnamefont
  {Busse}}, \bibinfo {author} {\bibfnamefont {Wolf-Christian}\ \bibnamefont
  {Müller}}, \bibinfo {author} {\bibfnamefont {Holger}\ \bibnamefont
  {Homann}}, \ and\ \bibinfo {author} {\bibfnamefont {Rainer}\ \bibnamefont
  {Grauer}},\ }\bibfield  {title} {\enquote {\bibinfo {title} {{Statistics of
  passive tracers in three-dimensional magnetohydrodynamic turbulence}},}\
  }\href {\doibase 10.1063/1.2818770} {\bibfield  {journal} {\bibinfo
  {journal} {Physics of Plasmas}\ }\textbf {\bibinfo {volume} {14}},\ \bibinfo
  {pages} {122303} (\bibinfo {year} {2007})},\ \Eprint
  {http://arxiv.org/abs/https://pubs.aip.org/aip/pop/article-pdf/doi/10.1063/1.2818770/16108097/122303\_1\_online.pdf}
  {https://pubs.aip.org/aip/pop/article-pdf/doi/10.1063/1.2818770/16108097/122303\_1\_online.pdf}
  \BibitemShut {NoStop}%
\bibitem [{\citenamefont {Verma}\ \emph {et~al.}(2020)\citenamefont {Verma},
  \citenamefont {Bhatnagar}, \citenamefont {Mitra},\ and\ \citenamefont
  {Pandit}}]{Akhilesh2020FPT}%
  \BibitemOpen
  \bibfield  {author} {\bibinfo {author} {\bibfnamefont {Akhilesh~Kumar}\
  \bibnamefont {Verma}}, \bibinfo {author} {\bibfnamefont {Akshay}\
  \bibnamefont {Bhatnagar}}, \bibinfo {author} {\bibfnamefont {Dhrubaditya}\
  \bibnamefont {Mitra}}, \ and\ \bibinfo {author} {\bibfnamefont {Rahul}\
  \bibnamefont {Pandit}},\ }\bibfield  {title} {\enquote {\bibinfo {title}
  {First-passage-time problem for tracers in turbulent flows applied to virus
  spreading},}\ }\href {\doibase 10.1103/PhysRevResearch.2.033239} {\bibfield
  {journal} {\bibinfo  {journal} {Phys. Rev. Res.}\ }\textbf {\bibinfo {volume}
  {2}},\ \bibinfo {pages} {033239} (\bibinfo {year} {2020})}\BibitemShut
  {NoStop}%
\bibitem [{\citenamefont {von Kameke}\ \emph {et~al.}(2011)\citenamefont {von
  Kameke}, \citenamefont {Huhn}, \citenamefont {Fern\'andez-Garc\'{\i}a},
  \citenamefont {Mu\~nuzuri},\ and\ \citenamefont
  {P\'erez-Mu\~nuzuri}}]{Kameke2011LagDispExp}%
  \BibitemOpen
  \bibfield  {author} {\bibinfo {author} {\bibfnamefont {A.}~\bibnamefont {von
  Kameke}}, \bibinfo {author} {\bibfnamefont {F.}~\bibnamefont {Huhn}},
  \bibinfo {author} {\bibfnamefont {G.}~\bibnamefont
  {Fern\'andez-Garc\'{\i}a}}, \bibinfo {author} {\bibfnamefont {A.~P.}\
  \bibnamefont {Mu\~nuzuri}}, \ and\ \bibinfo {author} {\bibfnamefont
  {V.}~\bibnamefont {P\'erez-Mu\~nuzuri}},\ }\bibfield  {title} {\enquote
  {\bibinfo {title} {Double cascade turbulence and richardson dispersion in a
  horizontal fluid flow induced by faraday waves},}\ }\href {\doibase
  10.1103/PhysRevLett.107.074502} {\bibfield  {journal} {\bibinfo  {journal}
  {Phys. Rev. Lett.}\ }\textbf {\bibinfo {volume} {107}},\ \bibinfo {pages}
  {074502} (\bibinfo {year} {2011})}\BibitemShut {NoStop}%
\bibitem [{\citenamefont {Von~Neumann}\ and\ \citenamefont
  {Richtmyer}(1950)}]{vonneumann1950method}%
  \BibitemOpen
  \bibfield  {author} {\bibinfo {author} {\bibfnamefont {J}~\bibnamefont
  {Von~Neumann}}\ and\ \bibinfo {author} {\bibfnamefont {R~D}\ \bibnamefont
  {Richtmyer}},\ }\bibfield  {title} {\enquote {\bibinfo {title} {A method for
  the numerical calculation of hydrodynamic shocks},}\ }\href@noop {}
  {\bibfield  {journal} {\bibinfo  {journal} {Journal of applied physics}\
  }\textbf {\bibinfo {volume} {21}},\ \bibinfo {pages} {232--237} (\bibinfo
  {year} {1950})}\BibitemShut {NoStop}%
\bibitem [{\citenamefont {Haugen}\ \emph {et~al.}(2022)\citenamefont {Haugen},
  \citenamefont {Brandenburg}, \citenamefont {Sandin},\ and\ \citenamefont
  {Mattsson}}]{Haugen2022ClusteringInertial}%
  \BibitemOpen
  \bibfield  {author} {\bibinfo {author} {\bibfnamefont {Nils E.~L.}\
  \bibnamefont {Haugen}}, \bibinfo {author} {\bibfnamefont {Axel}\ \bibnamefont
  {Brandenburg}}, \bibinfo {author} {\bibfnamefont {Christer}\ \bibnamefont
  {Sandin}}, \ and\ \bibinfo {author} {\bibfnamefont {Lars}\ \bibnamefont
  {Mattsson}},\ }\bibfield  {title} {\enquote {\bibinfo {title} {Spectral
  characterisation of inertial particle clustering in turbulence},}\ }\href
  {\doibase 10.1017/jfm.2021.1143} {\bibfield  {journal} {\bibinfo  {journal}
  {J. Fluid Mech.}\ }\textbf {\bibinfo {volume} {934}},\ \bibinfo {pages} {A37}
  (\bibinfo {year} {2022})}\BibitemShut {NoStop}%
\bibitem [{\citenamefont {Touchette}(2009)}]{touchette2009large}%
  \BibitemOpen
  \bibfield  {author} {\bibinfo {author} {\bibfnamefont {Hugo}\ \bibnamefont
  {Touchette}},\ }\bibfield  {title} {\enquote {\bibinfo {title} {The large
  deviation approach to statistical mechanics},}\ }\href@noop {} {\bibfield
  {journal} {\bibinfo  {journal} {Physics Reports}\ }\textbf {\bibinfo {volume}
  {478}},\ \bibinfo {pages} {1--69} (\bibinfo {year} {2009})}\BibitemShut
  {NoStop}%
\bibitem [{Note1()}]{Note1}%
  \BibitemOpen
  \bibinfo {note} {The rate function $I(s)$ must be convex up, hence in
  principle there must be at least a $s^4$ term but to the range of $s$ over
  which we have collapsed data this term is negligible.}\BibitemShut {Stop}%
\bibitem [{\citenamefont {Redner}(2001)}]{Redner2001guide}%
  \BibitemOpen
  \bibfield  {author} {\bibinfo {author} {\bibfnamefont {Sidney}\ \bibnamefont
  {Redner}},\ }\href@noop {} {\emph {\bibinfo {title} {A guide to first-passage
  processes}}}\ (\bibinfo  {publisher} {Cambridge university press},\ \bibinfo
  {year} {2001})\BibitemShut {NoStop}%
\bibitem [{Note2()}]{Note2}%
  \BibitemOpen
  \bibinfo {note} {The scaling range in Fig.~(\ref {fig:FPT}c) is not large
  enough to do a proper error analysis. A least-square fit to the data over a
  scaling range estimated by the eye gives an exponent of approximately
  $1.4$.}\BibitemShut {Stop}%
\bibitem [{\citenamefont {Cherstvy}\ and\ \citenamefont
  {Metzler}(2013)}]{Chertsvy2013HetDiff1}%
  \BibitemOpen
  \bibfield  {author} {\bibinfo {author} {\bibfnamefont {Andrey~G.}\
  \bibnamefont {Cherstvy}}\ and\ \bibinfo {author} {\bibfnamefont {Ralf}\
  \bibnamefont {Metzler}},\ }\bibfield  {title} {\enquote {\bibinfo {title}
  {Population splitting{,} trapping{,} and non-ergodicity in heterogeneous
  diffusion processes},}\ }\href {\doibase 10.1039/C3CP53056F} {\bibfield
  {journal} {\bibinfo  {journal} {Phys. Chem. Chem. Phys.}\ }\textbf {\bibinfo
  {volume} {15}},\ \bibinfo {pages} {20220--20235} (\bibinfo {year}
  {2013})}\BibitemShut {NoStop}%
\bibitem [{\citenamefont {Cherstvy}\ \emph {et~al.}(2013)\citenamefont
  {Cherstvy}, \citenamefont {Chechkin},\ and\ \citenamefont
  {Metzler}}]{Cherstvy2013HetDiff2}%
  \BibitemOpen
  \bibfield  {author} {\bibinfo {author} {\bibfnamefont {Andrey~G}\
  \bibnamefont {Cherstvy}}, \bibinfo {author} {\bibfnamefont {Aleksei~V}\
  \bibnamefont {Chechkin}}, \ and\ \bibinfo {author} {\bibfnamefont {Ralf}\
  \bibnamefont {Metzler}},\ }\bibfield  {title} {\enquote {\bibinfo {title}
  {Anomalous diffusion and ergodicity breaking in heterogeneous diffusion
  processes},}\ }\href {\doibase 10.1088/1367-2630/15/8/083039} {\bibfield
  {journal} {\bibinfo  {journal} {New Journal of Physics}\ }\textbf {\bibinfo
  {volume} {15}},\ \bibinfo {pages} {083039} (\bibinfo {year}
  {2013})}\BibitemShut {NoStop}%
\bibitem [{\citenamefont {Xu}\ \emph {et~al.}(2020)\citenamefont {Xu},
  \citenamefont {Liu}, \citenamefont {Li},\ and\ \citenamefont
  {Metzler}}]{Xu2020HetDiff}%
  \BibitemOpen
  \bibfield  {author} {\bibinfo {author} {\bibfnamefont {Yong}\ \bibnamefont
  {Xu}}, \bibinfo {author} {\bibfnamefont {Xuemei}\ \bibnamefont {Liu}},
  \bibinfo {author} {\bibfnamefont {Yongge}\ \bibnamefont {Li}}, \ and\
  \bibinfo {author} {\bibfnamefont {Ralf}\ \bibnamefont {Metzler}},\ }\bibfield
   {title} {\enquote {\bibinfo {title} {{Heterogeneous diffusion processes and
  nonergodicity with Gaussian colored noise in layered diffusivity
  landscapes}},}\ }\href {\doibase 10.1103/PhysRevE.102.062106} {\bibfield
  {journal} {\bibinfo  {journal} {Phys. Rev. E}\ }\textbf {\bibinfo {volume}
  {102}},\ \bibinfo {pages} {062106} (\bibinfo {year} {2020})}\BibitemShut
  {NoStop}%
\bibitem [{\citenamefont {Metzler}\ \emph {et~al.}(2014)\citenamefont
  {Metzler}, \citenamefont {Jeon}, \citenamefont {Cherstvy},\ and\
  \citenamefont {Barkai}}]{Metzler2014AnoDiff}%
  \BibitemOpen
  \bibfield  {author} {\bibinfo {author} {\bibfnamefont {Ralf}\ \bibnamefont
  {Metzler}}, \bibinfo {author} {\bibfnamefont {Jae-Hyung}\ \bibnamefont
  {Jeon}}, \bibinfo {author} {\bibfnamefont {Andrey~G.}\ \bibnamefont
  {Cherstvy}}, \ and\ \bibinfo {author} {\bibfnamefont {Eli}\ \bibnamefont
  {Barkai}},\ }\bibfield  {title} {\enquote {\bibinfo {title} {Anomalous
  diffusion models and their properties: non-stationarity{,} non-ergodicity{,}
  and ageing at the centenary of single particle tracking},}\ }\href {\doibase
  10.1039/C4CP03465A} {\bibfield  {journal} {\bibinfo  {journal} {Phys. Chem.
  Chem. Phys.}\ }\textbf {\bibinfo {volume} {16}},\ \bibinfo {pages}
  {24128--24164} (\bibinfo {year} {2014})}\BibitemShut {NoStop}%
\bibitem [{\citenamefont {Singh}\ \emph {et~al.}(2022)\citenamefont {Singh},
  \citenamefont {Mukherjee},\ and\ \citenamefont {Ray}}]{Samriddhi2022PRF}%
  \BibitemOpen
  \bibfield  {author} {\bibinfo {author} {\bibfnamefont {Rahul~K.}\
  \bibnamefont {Singh}}, \bibinfo {author} {\bibfnamefont {Siddhartha}\
  \bibnamefont {Mukherjee}}, \ and\ \bibinfo {author} {\bibfnamefont
  {Samriddhi~Sankar}\ \bibnamefont {Ray}},\ }\bibfield  {title} {\enquote
  {\bibinfo {title} {Lagrangian manifestation of anomalies in active
  turbulence},}\ }\href {\doibase 10.1103/PhysRevFluids.7.033101} {\bibfield
  {journal} {\bibinfo  {journal} {Phys. Rev. Fluids}\ }\textbf {\bibinfo
  {volume} {7}},\ \bibinfo {pages} {033101} (\bibinfo {year}
  {2022})}\BibitemShut {NoStop}%
\bibitem [{\citenamefont {Springel}\ and\ \citenamefont
  {Hernquist}(2003)}]{Volker2002Cosmo}%
  \BibitemOpen
  \bibfield  {author} {\bibinfo {author} {\bibfnamefont {Volker}\ \bibnamefont
  {Springel}}\ and\ \bibinfo {author} {\bibfnamefont {Lars}\ \bibnamefont
  {Hernquist}},\ }\bibfield  {title} {\enquote {\bibinfo {title} {Cosmological
  smoothed particle hydrodynamics simulations: a hybrid multiphase model for
  star formation},}\ }\href {\doibase 10.1046/j.1365-8711.2003.06206.x}
  {\bibfield  {journal} {\bibinfo  {journal} {Monthly Notices of the Royal
  Astronomical Society}\ }\textbf {\bibinfo {volume} {339}},\ \bibinfo {pages}
  {289--311} (\bibinfo {year} {2003})},\ \Eprint
  {http://arxiv.org/abs/https://academic.oup.com/mnras/article-pdf/339/2/289/3230545/339-2-289.pdf}
  {https://academic.oup.com/mnras/article-pdf/339/2/289/3230545/339-2-289.pdf}
  \BibitemShut {NoStop}%
\bibitem [{\citenamefont {Weinberger}\ and\ \citenamefont
  {Hernquist}(2022)}]{Weinberger2022Cosmo}%
  \BibitemOpen
  \bibfield  {author} {\bibinfo {author} {\bibfnamefont {R}~\bibnamefont
  {Weinberger}}\ and\ \bibinfo {author} {\bibfnamefont {L}~\bibnamefont
  {Hernquist}},\ }\bibfield  {title} {\enquote {\bibinfo {title} {Modelling
  multiphase gases in cosmological simulations using compressible multifluid
  hydrodynamics},}\ }\href {\doibase 10.1093/mnras/stac3708} {\bibfield
  {journal} {\bibinfo  {journal} {Monthly Notices of the Royal Astronomical
  Society}\ }\textbf {\bibinfo {volume} {519}},\ \bibinfo {pages} {3011--3026}
  (\bibinfo {year} {2022})},\ \Eprint
  {http://arxiv.org/abs/https://academic.oup.com/mnras/article-pdf/519/2/3011/48522828/stac3708.pdf}
  {https://academic.oup.com/mnras/article-pdf/519/2/3011/48522828/stac3708.pdf}
  \BibitemShut {NoStop}%
\bibitem [{\citenamefont {{Richard, Thibaud}}\ \emph
  {et~al.}(2022)\citenamefont {{Richard, Thibaud}}, \citenamefont {{Lesaffre,
  Pierre}}, \citenamefont {{Falgarone, Edith}},\ and\ \citenamefont {{Lehmann,
  Andrew}}}]{Thibaud2022Dissipation}%
  \BibitemOpen
  \bibfield  {author} {\bibinfo {author} {\bibnamefont {{Richard, Thibaud}}},
  \bibinfo {author} {\bibnamefont {{Lesaffre, Pierre}}}, \bibinfo {author}
  {\bibnamefont {{Falgarone, Edith}}}, \ and\ \bibinfo {author} {\bibnamefont
  {{Lehmann, Andrew}}},\ }\bibfield  {title} {\enquote {\bibinfo {title}
  {Probing the nature of dissipation in compressible mhd turbulence},}\ }\href
  {\doibase 10.1051/0004-6361/202142531} {\bibfield  {journal} {\bibinfo
  {journal} {Astron. Astrophys.}\ }\textbf {\bibinfo {volume} {664}},\ \bibinfo
  {pages} {A193} (\bibinfo {year} {2022})}\BibitemShut {NoStop}%
\bibitem [{\citenamefont {Hou}\ and\ \citenamefont
  {Li}(2007)}]{Hou2007ExponentialFilter}%
  \BibitemOpen
  \bibfield  {author} {\bibinfo {author} {\bibfnamefont {Thomas~Y.}\
  \bibnamefont {Hou}}\ and\ \bibinfo {author} {\bibfnamefont {Ruo}\
  \bibnamefont {Li}},\ }\bibfield  {title} {\enquote {\bibinfo {title}
  {Computing nearly singular solutions using pseudo-spectral methods},}\ }\href
  {\doibase https://doi.org/10.1016/j.jcp.2007.04.014} {\bibfield  {journal}
  {\bibinfo  {journal} {Journal of Computational Physics}\ }\textbf {\bibinfo
  {volume} {226}},\ \bibinfo {pages} {379--397} (\bibinfo {year}
  {2007})}\BibitemShut {NoStop}%
\bibitem [{\citenamefont {Higham}(2001)}]{higham2001algorithmic}%
  \BibitemOpen
  \bibfield  {author} {\bibinfo {author} {\bibfnamefont {Desmond~J}\
  \bibnamefont {Higham}},\ }\bibfield  {title} {\enquote {\bibinfo {title} {An
  algorithmic introduction to numerical simulation of stochastic differential
  equations},}\ }\href@noop {} {\bibfield  {journal} {\bibinfo  {journal} {SIAM
  review}\ }\textbf {\bibinfo {volume} {43}},\ \bibinfo {pages} {525--546}
  (\bibinfo {year} {2001})}\BibitemShut {NoStop}%
\bibitem [{\citenamefont {Cox}\ and\ \citenamefont
  {Matthews}(2002)}]{cox2002exponential}%
  \BibitemOpen
  \bibfield  {author} {\bibinfo {author} {\bibfnamefont {Steven~M}\
  \bibnamefont {Cox}}\ and\ \bibinfo {author} {\bibfnamefont {Paul~C}\
  \bibnamefont {Matthews}},\ }\bibfield  {title} {\enquote {\bibinfo {title}
  {Exponential time differencing for stiff systems},}\ }\href@noop {}
  {\bibfield  {journal} {\bibinfo  {journal} {Journal of Computational
  Physics}\ }\textbf {\bibinfo {volume} {176}},\ \bibinfo {pages} {430--455}
  (\bibinfo {year} {2002})}\BibitemShut {NoStop}%
\end{thebibliography}
% \onecolumngrid

\appendix

\section{Randomly-forced Burgers turbulence}
\label{app:Burgers}
\begin{figure*}
    \centering
    \includegraphics[width=0.9\textwidth]{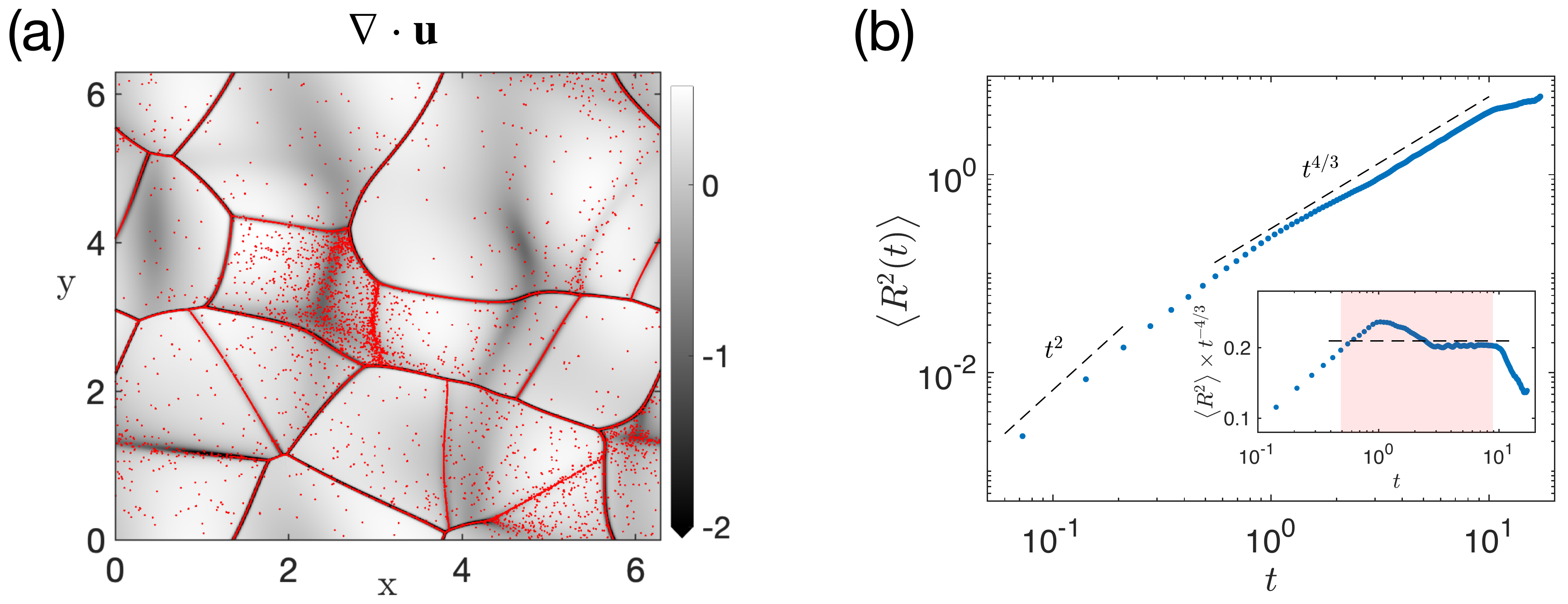}
    \caption{\textbf{Burgers turbulence.}
      (a) Snapshot of $\dive\uu$ in the non-equilibrium steady-state of the
      stochastically-forced 2D Burgers equations; shocks are exhibited as
      dark filaments where $\dive\uu\ll 0$.
      The plot is overlaid with the instantaneous tracer positions denoted
      by the red dots; the tracers clearly cluster at the shocks. 
      (b) Log-log plots of the mean-squared displacements of the tracers,
      $\MSD(t)$, versus $t$.
      Superdiffusive transport: $\MSD \sim t^{\gamma}$ with
      $\gamma\approx 4/3$, at late times, $t>1$,
      is clearly visible ($t$ has been normalized by the eddy-turnover time,
      $\TE$). 
      The inset shows the plot of $\MSD$ compensated by $t^{-4/3}$ with the
      shaded area denoting the scaling regime.
      Local slope analysis across this regime results in the mean-squared
      displacement exponent, $\gamma=1.30\pm0.08$.}
    \label{fig:Burg}
\end{figure*}

We investigate the randomly-forced 2D Burgers equations which models a
pressure-less, irrotational flow ($\Psi=0$):
\begin{equation}
    \frac{\partial}{\partial t}\uu +  \uu\cdot\nabla \uu = \nu\nabla^2\uu + \ffext\,,
    \label{eq:Burgers}
\end{equation}
where $\ffext$ is an external force. 
We have chosen $\ffext$ to be statistically identical to that in
$\RC$ (i.e. irrotational, white-in-time and restricted to the same range of
wavenumbers) such that $\dive\uu=0$ at all times. 
A snapshot of $\dive\uu$ in the non-equilibrium steady state of Burgers
turbulence, displayed in \subfig{fig:Burg}{a}, clearly exhibits shocks as
long filamentary structures characterized by $\dive\uu < 0$. 
At initial time, $t=0$, the tracers are uniformly distributed in the domain.
As they evolve with the flow, they are drawn into the shocks. 
Due to the absence of the pressure-gradient force, they remain trapped in the
shocks forever, as seen in \subfig{fig:Burg}{a} where the tracers, denoted as
red dots, straddle the shocks. 

\subfig{fig:Burg}{b} displays a log-log plot of the mean-squared displacement,
$\MSD$, of tracers versus $t$. 
Beyond the ballistic scaling regime at early times, $t\ll1$, there is a
transition to superdiffusion at $t\sim\mathcal{O}(1)$. 
The superdiffusive regime extends over a decade to late times $t\gg1$ before
there is an under-sampling of distinct tracer trajectories due to the
absorption of particles in the shocks. 
Within the regime of superdiffusion,
$\MSD\sim t^\gamma$ with $\gamma\approx 4/3$ --
same as what we also found in $\RC$ and $\RMo$. 
The plot of $\MSD$ compensated by $t^{-4/3}$, show in the inset of
\subfig{fig:Burg}{b}, is also roughly horizontal in the superdiffusive
regime marked by the shaded area. 
Local slope analysis of the log-log graph across the superdiffusive regime
yields $\gamma=1.30\pm0.08$ which is within errorbars of the exponents
extracted from $\RC$ and $\RMo$. 

This shows that the superdiffusive transport of tracers which we uncover in
predominantly irrotational, governed by the
compressible Navier-Stokes equations, is observed also in forced
Burgers equation, even with the same exponent. 
%==========================%
\section{Direct Numerical simulation}
\label{sec:numsim}
The code we use for direct numerical simulations (DNS) of the
compressible, isothermal, Navier Stokes equations in two dimensions (2D)
has been used in our earlier
work~\citep[][Supplementary Material]{de2026intermittency}.
For the sake of completeness we repeat it here. 

The governing equations are as follows:
\begin{widetext}
\begin{subequations}
\begin{align}
    & \frac{\partial\rho}{\partial t} + \grad\cdot(\rho\uu)=0\,,
    \label{eq:Cont1app}
    \\
    & \frac{\partial\uu}{\partial t} + \uu\cdot\grad\uu =  [2\nu{\bm S}+{\bm I}_2(\zbulk\dive\uu-\cs^2)]\cdot\grad(\ln\rho) +
    \nu\grad^2\uu + \dive(\zbulk\dive\uu) +  \ff_{\rm ext}\,.
    \label{eq:Mom1app}
\end{align}
\end{subequations}
\end{widetext}
Here $\uu(\xx,t)$ is the velocity field,
$\rho(\xx,t)$ is the density field,
$P$ is the pressure,
$\nu$ is the kinematic shear viscosity
taken to be a constant~\citep{Haugen2022ClusteringInertial,
  Federrath2011Machno, Thibaud2022Dissipation}.
$\zbulk$ is the bulk viscosity,
$\delta_{ij}$ is the Kronecker delta,
$S_{ij} \equiv (1/2)\left(\delj \ui + \deli \uj\right) -
           (1/2)\delta_{ij}\delk\uk$ 
is the traceless rate-of-strain tensor,
and the equation of state is given by 
$P = \cs^2\rho$, where $\cs$ is the (constant) speed of sound. 
Here, $f_{{\rm ext}\,,i}=\mathcal{P}_{ij}f_j$ and
${\bm I}_2$ is the $2\times2$ identity matrix.
In 2D, Eq.~\eqref{eq:Mom1app} can be rewritten in terms of two scalars -- the
vorticity, $\omega=\zhat\cdot\curl\uu$, and the divergence of velocity,
$\mathcal{D}=\dive\uu$.

We use a pseudo-spectral code to solve \eq{eq:Mom1app} rewritten
in $\omega$ and $\mD$. 

\subsection{De-aliasing}
\label{sec:dealias}
To remove aliasing errors, we use an exponential
filter~\citep{Hou2007ExponentialFilter}:
\begin{equation}
    \mathcal{F}(k) = \exp\left[-36.0\left(\frac{k}{\km}\right)^{36.0}\right]\,,
    \label{eq:filter}
\end{equation}
where $\km=N/2$ is the highest allowed wave-number in the simulation,
$N$ the number of grid points in a given dimension. 
In our simulations, $N=2048$.
The filter $\mathcal{F}(k)$ is approximately unity up to $k\approx N/3$ and
drops sharply but smoothly to almost zero beyond that, essentially all modes
with $k>N/3$ are strongly damped.
It is a better choice than the Galerkin-based discrete cut-off filter
which are usually used for incompressible turbulence because it reduces
oscillations around shocks which result from
Galerkin-trucation, enabling us to simulate supersonic turbulence at
high Mach numbers while maintaining relatively large Reynolds numbers.

\subsection{External force}
\label{sec:force}
The quantity, $\fvec(\xvec,t)$, is a white-in-time random noise. 
Its spectrum in the spatial domain is restricted to large spatial scales
(small $k$). 
In our simulations, we use
\begin{subequations}
  \begin{align}
    \widehat{\fvec}_{k_x}(\kvec,t_n) &= A  \cos(\theta_{\kvec,n})e^{i\phi_{\kvec,n}}\/,\  \\
\text{and}\quad    \widehat{\fvec}_{k_y}(\kvec,t_n) &= A \sin(\theta_{\kvec,n})e^{i\phi_{\kvec,n}}\/,
  \end{align}
\end{subequations}
\label{eq:cNSForceSim}
where the amplitude,
\begin{equation}
    A = \begin{cases}
    f_0\sqrt{\delta t} &\text{for}\quad k=\kf \,; \\
    0 &\text{otherwise}\,.
    \end{cases}
    \label{eq:Forc_amp}
\end{equation}
$f_0$ is constant, and $\theta_{\kvec,n}$ and $\phi_{\kvec,n}$ are uniformly
distributed random numbers belonging to the interval $[0,2\pi]$. 
We choose $\kf=3$. 
The ratio of power being injected into the solenoidal and irrotational modes
is tuned by applying the following projection operator,
\begin{equation}
    \hat{\mathcal{P}}_{ij}=\xi\delta_{ij}+ (1-2\xi)k_ik_j/k^2\,,
    \label{eq:Pij_cNS}
\end{equation}
which yields $\fvec_{\rm ext}(\xx,t)$.
\subsection{The shock viscosity $\zbulk$}
\label{sec:shock}
The shock viscosity, $\zbulk$, is a spatially non-uniform bulk viscosity which
is locally enhanced in the neighbourhood of shocks and serves to resolve the
strong shocks on a
finite-resolution grid in the high Mach number simulations. 
We evaluate $\zbulk$ on the basis of the prescription outlined in
Ref.~\cite{vonneumann1950method}. 
At each grid $(i,j)\equiv\xvec$,
\begin{equation}
    \zbulk(\xvec) = c_{\rm shock}\avg{\overline{\mathcal{D}}}_5\delta x \delta y\,,
    \label{eq:shockvis}
\end{equation}
where $\delta x = \delta y=L/N$ are the grid-spacings along each direction, and $\avg{...}_5$ denotes the average over a $5\times5$ domain centred at $(i,j)$.
\begin{equation}
    \overline{\mathcal{D}} = \begin{cases}
    |\dive\uu| &\text{for}\quad \dive\uu<0 \,; \\
    0 &\text{otherwise}\,.
    \end{cases}
    \label{eq:Dbar}
\end{equation}
We choose $c_{\rm shock}\sim\mathcal{O}(1)$. 
Thus, $\zbulk$ is non-zero around shocks, decreasing quickly with distance from them, and zero elsewhere. 

\subsection{Timestepping}
\label{sec:time}
We implement the white-in-time external force with Euler Marayuma
scheme~\citep{higham2001algorithmic} 
and exponential Adams-Bashforth method~\citep{cox2002exponential} for
timestepping the rest of the equation. 
%-------------------------------------------
\subsection{Lagrangian particle tracking}
The position of the $\rm{m}$-th tracer, $\XXm(t)$,
evolves according to
\begin{subequations}
  \begin{align}
\frac{d\XXm(t)}{dt}&=\Vvec^{\rm m}(t) \\
\text{where}\quad \Vvec^{\rm m}(t)&=\uu(\xx,t)\delta(\xx-\XXm)\/.
\end{align}
\end{subequations}
The velocity at the off-grid position of a tracer is obtained by linear
interpolation from the neighboring grid points and the timestepping
is done by a Runge-Kutta scheme. 
%--------------------------------
\subsubsection{Evaluation of CPDF}
We evaluate all CPDfs by the rank-order method to avoid binning errors. 
In this method, we arrange all the $M$ values of a variable in descending
order, assigning each a rank from 1 to $M$. 
On the y-axes, we we plot the ranks, $Q$, corresponding to the values on
the x-axis.
%----------------------------
\section{Results}
%-------------------------------
\begin{table}
    \centering
    \begin{tabular}{|c|c|c|c|c|c|c|}
    \hline
    Run & $\Ma$ & $\Rey$ & $\LI$ & $\TE$ & $\Psi$
    \\[3pt]
    \hline
    \hline
    \RC & $1.15$ & $4\times10^3$ & $1.10$ & $48$ & $0.24\pm0.06$ \\
    \hline
    \RMo & $1.24$ & $4.4\times10^3$ & $1.10$ & $46$ & $0.35\pm0.05$ \\
    \hline
    \RMt & $1.10$ & $4\times10^3$ & $1.12$ & $51$ & $0.46\pm0.08$ \\
    \hline
    \RMth & $1.20$ & $4.3\times10^3$ & $1.23$ & $53$ & $0.61\pm0.04$ \\
    \hline
    \RS & $1.10$ & $4\times10^3$ & $1.57$ & $74$ & $0.74\pm0.03$ \\
    % &  &  &  &  &  &  & \\
    % R6 & $1$ & $0.17$ & $373$ & $3.2\times10^3$ & $1.29$ & $74$ & $0.97\pm0.01$ \\ 
    \hline
    \end{tabular}
    \caption{\small \textbf{Parameters characterizing the turbulent steady-state} of Eq.(1) in the main text. 
    $Ma=\urms/\cs$ is the turbulent Mach number, where $\urms$ is the
    root-mean-square velocity of the flow,
      $\Rey=\frac{\urms L}{\nu}$ is the box size Reynolds number, $\LI=\pi\sum_{k}[E(k)/k]\Big/\sum_{k}E(k)$ is the integral length scale, $E(k)$ is the kinetic energy spectrum, $\TE=\LI/\urms$ is the large-eddy turnover time, and $\Psi$ is the solenoidal ratio. We choose $dt=10^{-3}$ $\cs=0.02$ and
      $\nu=3.5\times10^{-5}$.}
    \label{tab:para}
\end{table}
%-------------------------------

\begin{figure}
    \centering
    \includegraphics[width=0.8\columnwidth]{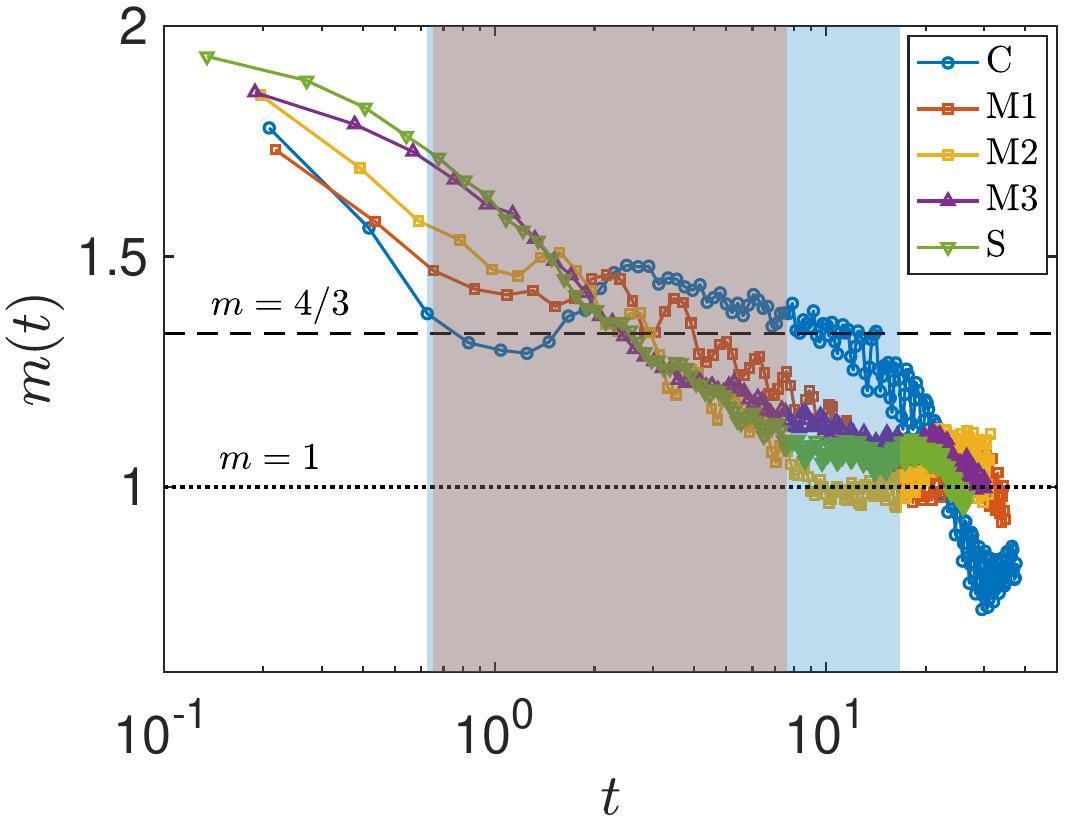}
    \caption{Local slopes, $m(t)$, of the log-log plots of $\MSD$ versus $t$
      for different runs shown in Fig. 1(b) in the main text;
      the blue and brown shaded regions denote the regimes of superdiffusion
      in \RC and \RMo, respectively, across which the mean and standard
      deviation of $m(t)$ yield the respective values of $\gamma$ and their
      errorbars.}
    \label{fig:locslop}
\end{figure}

The key parameters of our DNS are summarized in Table~\ref{tab:para}.
Local slopes of the log-log plots of $\MSD$ versus $t$,
shown in Fig. 1(b) of the main text, are plotted in
Fig.~\ref{fig:locslop}.

\begin{figure}
    \centering
    \includegraphics[width=0.8\columnwidth]{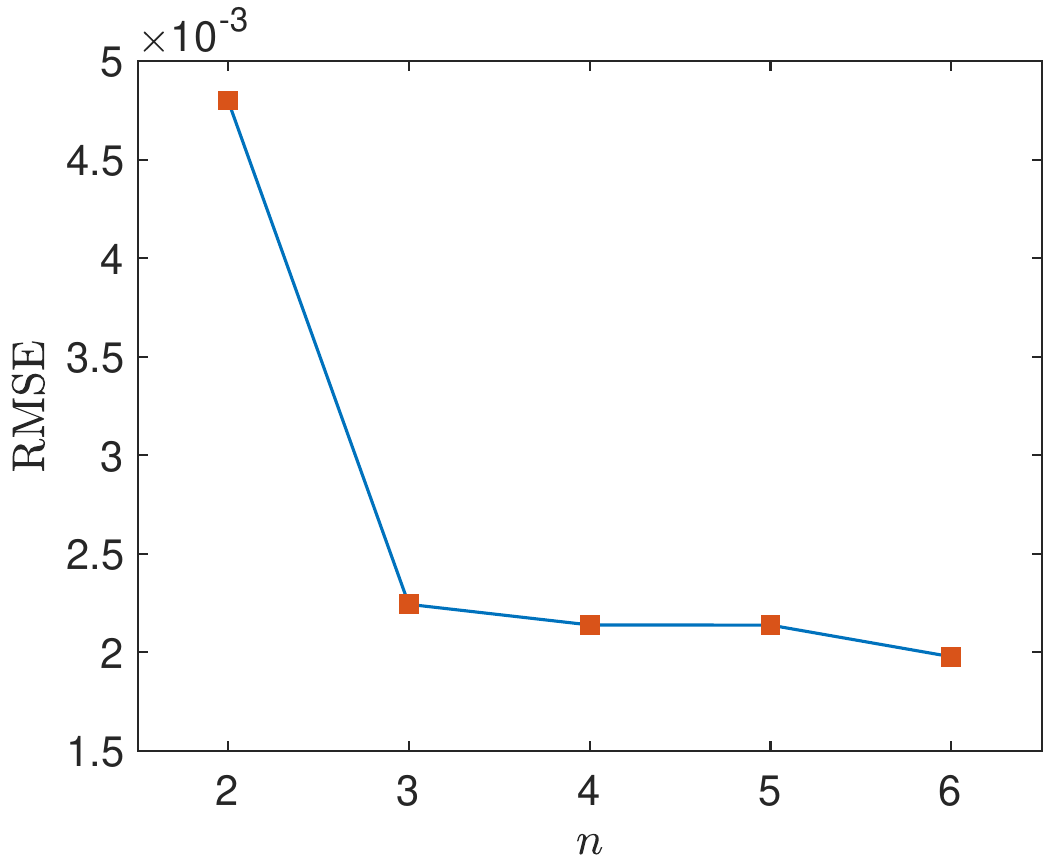}
    \caption{Root-mean-squared error (RMSE) of the nonlinear least-squares
      (NLS) fit to rate functoin $I(s)$ for $T\approx120$ in \RC, shown in
      Fig. 2 in the main text, versus the degree, $n$, of the
      fitting polynomials of the form, $f(x)=\sum_{i=2}^n a_ix^i$. 
    The RMSE converges for polynomials of degree $n\geq3$.}
    \label{fig:nlsrmse}
\end{figure}

The error in least square fit of a polynomial to  $I(s)$ in
$\RC$ (Fig. 2 in main text) as function of the degree $n$ of the
fitting polynomial is shown in Fig.~\ref{fig:nlsrmse}.

\begin{figure*}
    \centering
    \includegraphics[width=0.9\textwidth]{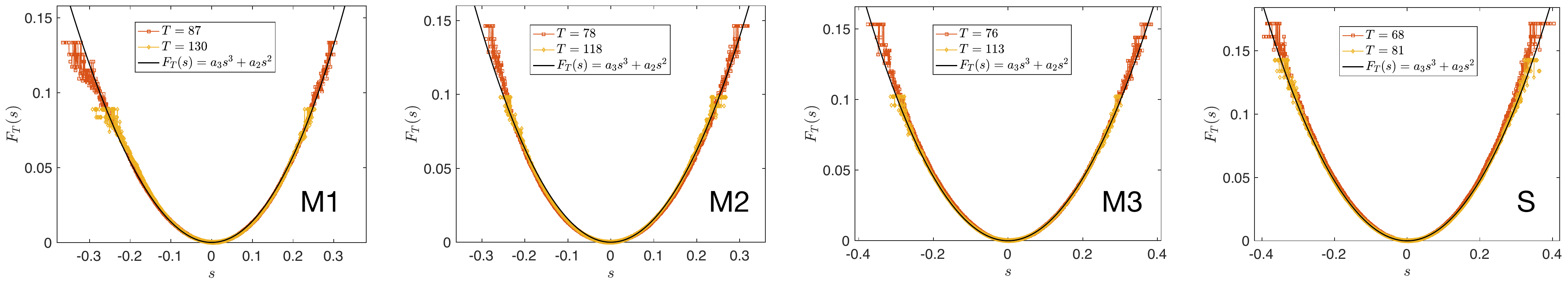}
    \caption{ Plot of the rate function $I(s)$, defined in \eq{eq:LDP},
      versus $s$ for all the runs except \RC.
      The plot is obtained by plotting the LHS of \eq{eq:LDP}
      for two large values of  $T$.
      The data collapses on a $T$-independent curve. 
      The black curve is the least-square fit to the data using
      the polynomial $\atwo s^2 + \ath s^3$.
      \label{fig:LDPrest}
    }
\end{figure*}

The PDF of $R_x/T$ for several large values of $T$
for the all the runs (except \RC ) are displayed in
\Fig{fig:LDPrest}.
In all of these cases we find that  
the left hand side of \eq{eq:LDP} converges to a
$T$-independent  function $I(s)$ which is plotted
 as the black line.

The Lagrangian velocity autocorrelation function is defined as 
\begin{equation}
\CL(\tau)=\avg{\Vvec^{\rm m}(0+\tau)\cdot\Vvec^{\rm m}(0)}
\end{equation}
where $\avg{\cdot}$ denotes averaging over different values of ${\rm m}$ and
uncorrelated snapshots, at  $t=0$.

%---------------------------

\begin{figure}
    \centering
    \includegraphics[width=0.8\linewidth]{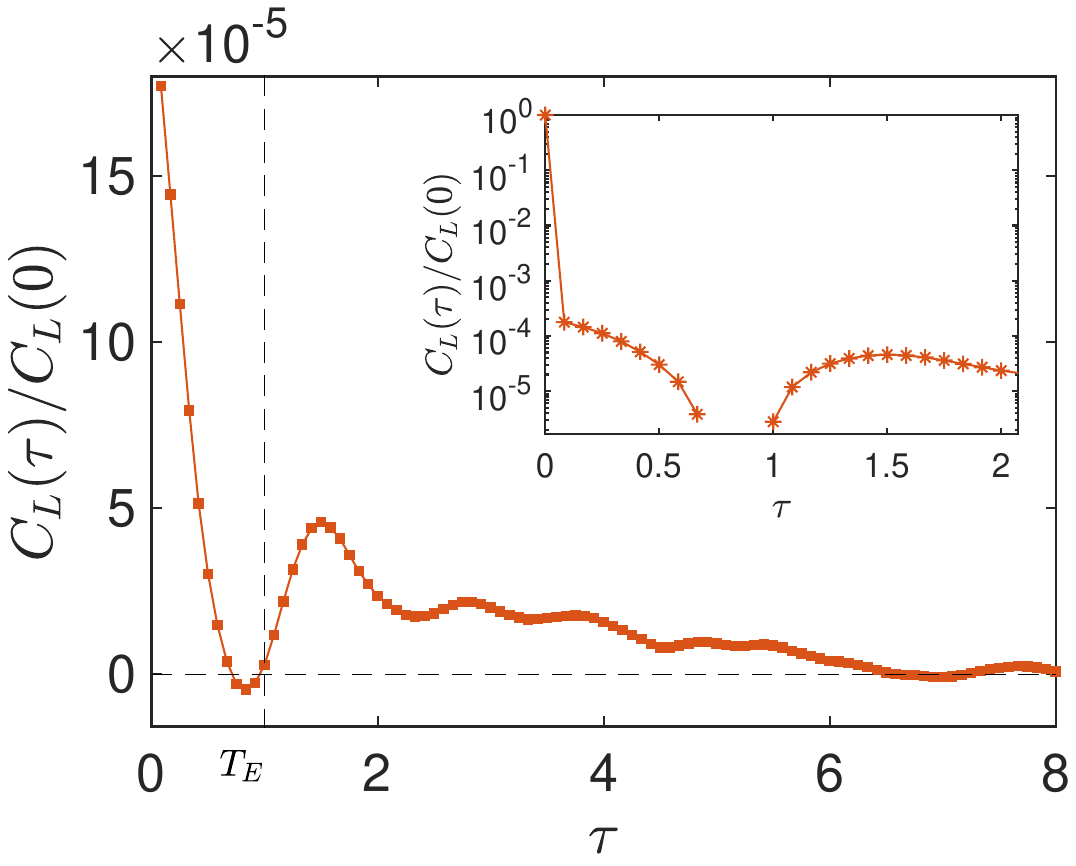}
    \caption{\textbf{Velocity autocorrelation function.} (a) Lagrangian
      velocity autocorrelation function for  $\RC$ of cNS turbulence.
      The inset shows the same plot in lin-log scale.
      The auto-correlation function is close to zero
      for $\tau \sim 0.1\TE$. 
    \label{fig:VelCorrfn} }
\end{figure}

\end{document}